\documentclass{amsart}[10pt]

\usepackage{amscd,amssymb,latexsym,url,verbatim,algorithmic,graphicx,color}

\usepackage[dvips]{epsfig}
\usepackage{tikz,tikz-cd}
\usetikzlibrary{decorations.pathmorphing}

\usepackage{cases,amsmath}

\title {Weak symmetry breaking and abstract simplex paths}

\author{Dmitry N. Kozlov}

\address{Department of Mathematics, University of Bremen, 28334
  Bremen, Federal Republic of Germany}

\email{dfk@math.uni-bremen.de}
\newtheorem{theorem}{Theorem}[section]
\newtheorem{df}[theorem]{Definition}
\newtheorem{thm}[theorem]{Theorem} 

\newtheorem{prop}[theorem]{Proposition}
\newtheorem{lm}[theorem]{Lemma}

\newtheorem{example}[theorem]{Example}

\newcommand{\nin}{\noindent}
\newcommand{\pr}{\nin{\bf Proof.} }

\newcommand{\cs}{{\mathcal S}}
\newcommand{\inte}{\text{\rm int}\,}
\newcommand{\da}{\Delta}
\newcommand{\Div}{\text{\rm Div}\,}

\newcommand{\supp}{\text{\rm supp}\,}
\newcommand{\op}{\text{\rm op}\,}

\newcommand{\wti}{\widetilde}
\newcommand{\rr}{{\mathbb R}}

\newcommand{\ra}{\rightarrow}

\newcommand{\last}{\rm{last}}

\newcommand{\dar}{\downarrow}
\newcommand{\dara}{\dar & & }
\newcommand{\darb}{&\dar &}
\newcommand{\darc}{&&\dar}
\newcommand{\dard}{&&&\dar}
\newcommand{\mc}{\multicolumn}
\newcommand{\two}[2]{\hline\mc{1}{|c}{#1} & \mc{1}{c|}{#2} \\ \hline}
\newcommand{\three}[3]{\hline\mc{1}{|c}{#1} & #2 & \mc{1}{c|}{#3} \\ \hline}
\newcommand{\four}[4]{\hline\mc{1}{|c}{#1} & #2 & #3 & \mc{1}{c|}{#4} \\ \hline}
\newcommand{\five}[5]{\hline\mc{1}{|c}{#1} & #2 & #3 & #4 & \mc{1}{c|}{#5} \\ \hline}
\newcommand{\cuben}{\textrm{Cube}_{[n]}}
\newcommand{\sm}{\setminus}

\numberwithin{equation}{section}
\numberwithin{figure}{section}
\numberwithin{table}{section}

\begin{document}

\begin{abstract}
Motivated by questions in theoretical distributed computing, we
develop the combinatorial theory of abstract simplex path
subdivisions. Our main application is a~short and structural proof of
a~theorem of Casta\~neda \& Rajsbaum. This theorem in turn implies
the solvability of the weak symmetry breaking task in the immediate
snapshot wait-free model in the case when the number of processes is
not a~power of a~prime number.
\end{abstract}

\maketitle

\section{Introduction}
The mathematical research presented here is motivated by problems of
theoretical distributed computing. Even though, the computer science
background is, strictly speaking, not necessary to read the
definitions and the proofs in this paper, we feel it is beneficial to
review the broader context, before delving into the description of our
results. The reader who is only interested in the mathematical part,
may skip our explanations, and proceed directly to
Section~\ref{sect:3} at this point.

In theoretical distributed computing one studies solvability of
standard tasks under various computational models. The computational
model which we consider here is the following. We have $n+1$ {\it
  asynchronous} processes, which communicate with each other using the
read/write shared memory, which in turn consists of atomic registers
assigned to individual processes. The operations allowed for each
process are \emph{write} and \emph{snapshot read}. The operation
\emph{write} writes whatever value the process wants into its register
in the shared memory. The operation \emph{snapshot read} reads the
entire shared memory atomically.

Each process executes a~wait-free protocol; with crash failures
allowed.  A~\emph{crash failure} means that the process which failed
simply stops executing its protocol, rather than, say, sending the
wrong information. The protocol is called \emph{wait-free}, if,
intuitively speaking, the processes are not allowed to \emph{wait} for
each other, i.e., to make their executions contingent on hearing from
other processes. In practical terms, this means that when a~process
$A$ did not hear from the process $B$ for a~while, and now needs to
decide on an~output value, its decision must be such that the total
produced output will be valid, no matter whether the process $B$
crashed, or whether it will suddenly revive its execution once $A$ has
chosen its output. Due to asynchrony, the process $A$ cannot
distinguish between these two options at this point of the execution.

For brevity, we refer to this entire computational model as
\emph{read/write wait-free} model. There is a~large class of
computational models, which have been proved to be equivalent to each
other. Since this class includes the read/write wait-free model, all
the statements pertaining to solvability of certain tasks proved for
this model are actually quite general, and say something about the
entire subject of wait-free asynchronous computation. We refer the
interested reader to \cite{HKR,HS,He91,AW} for further background.

Once the computational model is fixed, it becomes interesting to
understand which tasks are solvable in this model. To gain structural
insight, one concentrates on the questions of solvability of the
so-called \emph{standard tasks}. The standard task which is central to
this paper is the so-called \emph{weak symmetry breaking task}, which
will be abbreviated to WSB. In this task, the processes have no input
values, just their ID's. Their output values are~$0$ and~$1$. The task
is solvable if there exists an read/write wait-free protocol such that
in every execution in which all processes decide on an output value,
not all processes decide on the same one. In addition, this protocol
is required to be \emph{rank-symmetric} in the sense which we now
describe.

Let $I_1,I_2\subseteq\{0,\dots,n\}$, such that $|I_1|=|I_2|$, and let
$r:I_1\to I_2$ be the~order-preserving bijection. Assume $E_1$ is an
execution of the protocol, in which the set of participating processes
is $I_1$, and $E_2$ is an execution of the protocol, in which the set
of participating processes is $I_2$. Then, the protocol is called
rank-symmetric if for every $i\in I_1$, the process with ID~$i$ must
decide on the same value as the process with ID~$r(i)$. The protocol
will certainly be rank-symmetric, if each process only compares its ID
to the ID's of the other participating processes, and makes the final
decision based on the relative rank of its~ID.

Another important task is \emph{renaming}. The renaming task,
see~\cite{Attiya}, is defined as follows: the $n+1$ processes here
have unique input names, taken from some large universe of inputs, and
need to decide on unique output names from the set
$\{0,\dots,K\}$. The processes also have process ID's labeled
$0,\dots,n$, and it is requested for the protocol to be
\emph{anonymous}. The protocol is called anonymous if its
execution does not depend on the process ID, only the process input
value, and whatever information the process receives during the
execution. In other words, two different processes should decide on an
identical output value if they receive same inputs and same
information from the outside world, even though they will have
different~ID's. It has been proved that for $n+1$ processes the WSB is
equivalent to $(2n-1)$-renaming in the read/write wait-free model. So deciding
whether WSB is solvable will also tell us whether $(2n-1)$-renaming
for $n+1$ processes is solvable. We refer the reader to
\cite{AW,HKR,Ga06} for further information on the renaming task.

\section{Using simplex paths to construct a protocol}

\subsection{Solvability of the Weak Symmetry Breaking} $\,$

\nin
At present time, the solvability of WSB in the read/write wait-free model is
completely understood. For some time, it was believed that WSB is not
solvable for any number of processes, However, this turned out to be
incorrect. The correct answer is: 
\begin{quote}
\emph{Weak Symmetry Breaking for $n+1$ processes is solvable in
  read/write wait-free model if and only if $n+1$ is not a~prime
  power.}
\end{quote}  
While impossibility of WSB when $n+1$ is a~prime power has been known
for a~while, the solvability of WSB when $n+1$ is \emph{not} a~prime
power (the smallest example here is clearly when we have $6$
processes) has involved a~few turns in the literature. The final point
in the matter was put by Casta\~neda \& Rajsbaum, \cite{CR2}, see also
\cite{Attiya} for the prehistory. The important and interesting paper
of Casta\~neda \& Rajsbaum serves as a~motivation and the entry point
for the research presented in this article. We recommend that the
reader acquaints himself with the contents of~\cite{CR2}.

There is a~way of reducing the computability of WSB in read/write
wait-free model to a~purely mathematical existence question. It
involves the Anonymous Computability Theorem of Herlihy \& Shavit,
\cite{HS}, so its complete description would be too technical for the
current presentation. Therefore, for the sake of brevity, we opt to
confine ourselves to presenting a~mathematical formulation directly,
and then refering the reader, who is interested in the better
understanding of the connection between distributed computing and the
topological context to the above mentioned work of Casta\~neda \&
Rajsbaum, \cite{CR2}, as well as to~\cite{HKR}. The necessary background
in combinatorial topology can be found in~\cite{book}, while more on the
topology of protocol complexes can be found in~\cite{subd,view}. In the remainder 
of this section we give an equivalent reformulation of the solvability of WSB
in read/write wait-free model in terms of simplicial subdivisions.

\subsection{Notations} $\,$

\nin
First, we introduce some terminology.  We let $[n]$ denote the set
$\{0,\dots,n\}$, in particular, $[1]=\{0,1\}$ will denote the set of
boolean values. For a boolean value $c$ we let $\bar c$ denote the
negation of~$c$. For every $n\geq 1$ we let $\da^n$ denote the
so-called standard $n$-simplex. This is a~geometric simplicial complex,
which has a~unique $n$-simplex, spanned by the unit coordinate vectors
in $\rr^{n+1}$. The vertices of $\da^n$ are indexed by the elements of
the set $[n]$. More generally, the $k$-dimensional boundary simplices
of $\da^n$ are indexed by the subsets $I\subseteq[n]$, such that
$|I|=k+1$; we shall denote such a boundary simplex $\da^I$ and
identify its set of vertices with~$I$. For every two subsets
$I,J\subseteq[n]$, such that $|I|=|J|$, there exists a~unique
order-preserving bijection $r_{I,J}:I\ra J$. This order-preserving
bijection induces a~linear isomorphism $\varphi_{I,J}:\da^I\ra\da^J$.

A~finite geometric simplicial complex $\Div(\da^n)$ is called
a~\emph{subdivison} of $\da^n$ if every simplex of $\Div(\da^n)$ is
contained in a~simplex of $\da^n$, and every simplex of $\da^n$ is the
union of those simplices of $\Div(\da^n)$ which it contains. Clearly,
for every $I\subseteq[n]$, we have an induced subdivision
$\Div(\da^I)$.  For each simplex $\sigma\in\Div(\da^n)$, we let
$\supp(\sigma)$ denote the unique simplex $\tau$ of $\da^n$ such that
the interior of $\sigma$ is contained in the interior of~$\tau$.
Finally, for an arbitrary simplicial complex $K$, we let $V(K)$ denote
the set of vertices of~$K$. 

\subsection{Hereditary subdivisions and compliant labelings} $\,$

\nin
The next definition is crucial for the simplicial approach to the
rank-symmetric protocols.

\begin{df}
 A subdivision $\Div(\da^n)$ of the standard $n$-simplex $\da^n$ is
 called {\bf hereditary} if for all $I,J\subseteq[n]$ such that
 $|I|=|J|$, the linear isomorphism $\varphi_{I,J}$ is also an
 isomorphism of the subdivision restrictions $\Div(\da^I)$ and
 $\Div(\da^J)$.

A labeling $\lambda:V(\Div(\da^n))\ra L$, where $L$ is an~arbitrary
set of labels, is called {\bf compliant} if for all $I,J\subseteq[n]$
such that $|I|=|J|$, and all vertices $v\in V(\Div(\da^I))$, we have
\begin{equation}
\lambda(\varphi_{I,J}(v))=\lambda(v).
\end{equation}
\end{df}

\begin{figure}[hbt]
\begin{center}

  \input{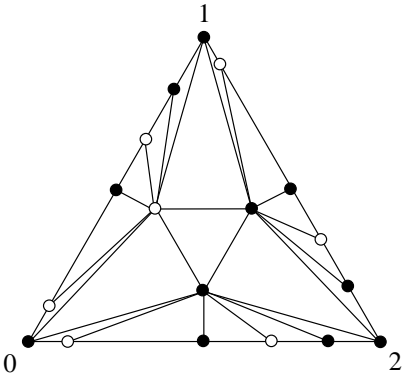}  

\end{center}
\caption{A hereditary subdivision with a~compliant binary labeling.}
\label{fig:chr1}
\end{figure}

An example of a~hereditary subdivision with a compliant binary
labeling is shown in Figure~\ref{fig:chr1}.

\begin{df}
Given a~subdivision $\Div(\da^n)$, the labeling
$\chi:V(\Div(\da^n))\ra[n]$ is called a~{\bf coloring} if
\begin{enumerate}
\item[(1)] any two vertices of $\Div(\da^n)$, which are connected by
  an edge, receive different colors;
\item[(2)] for any $v\in V(\Div(\da^n))$, we have $\chi(v)\in V(\supp(v))$.
\end{enumerate}
A~subdivision which allows a~coloring is called {\bf chromatic}.
\end{df}

Clearly, not every subdivision is chromatic, however, if such
a~coloring exists, then it must be unique. The subdivision shown in
Figure~\ref{fig:chr1} is chromatic.

The next theorem provides the key bridge between the solvability
of the WSB as a~distributed computing task, and the mathematics
of compliant labelings of hereditary chromatic subdivisions.

\begin{thm} \label{thm:equiv}
The task WSB for $n+1$ processes is solvable in read/write wait-free model if
and only if there exists a~hereditary chromatic subdivision
$\Div(\da^n)$ of~$\da^n$ and a~binary compliant labeling
$b:V(\Div(\da^n))\ra\{0,1\}$, such that for every $n$-simplex
$\sigma\in\Div(\da^n)$ the restriction map $b:V(\sigma)\ra\{0,1\}$ is
surjective.
\end{thm}

In general, if all vertices of an~$n$-simplex $\sigma\in\Div(\da^n)$
get the same label under $b$, one calls $\sigma$ \emph{monochromatic},
so the surjectivity condition is equivalent to saying that there are
no monochromatic $n$-simplices under~$b$. We shall also talk of
$0$-monochromatic and $1$-monochromatic $n$-simplices when we need to
specify the label assigned to the vertices of~$\sigma$. The
Theorem~\ref{thm:equiv} is due to the work of Herlihy \& Shavit. As we
said earlier, we assume its proof for the purposes of this paper, and
proceed with a purely mathematical analysis of the conditions.

In \cite{CR2}, Casta\~neda \& Rajsbaum proved that if $n+1$ is not
a~prime power, there exists a~hereditary chromatic subdivision
$\Div(\da^n)$ together with a~binary compliant labeling
$b:V(\Div(\da^n))\ra\{0,1\}$, satisfying conditions of
Theorem~\ref{thm:equiv}. This settles the solvability of WSB in read/write
wait-free model when the number of processes is not a~prime power.
The idea of the proof of Casta\~neda \& Rajsbaum is as follows:
\begin{itemize} 
\item start with some $\Div(\da^n)$ and some labeling $b$, which may
  have monochromatic simplices;
\item connect the monochromatic simplices if possible by simplicial
  paths of even lengths;
\item subdivide each path further, so as to eliminate the two
  monochromatic end $n$-simplices.
\end{itemize}

\subsection{Geometric simplex paths} $\,$

\nin
The hard part, and the crux of the proof in \cite{CR2} lies in
a~sophisticated and ingenious algorithm for the simplicial path
subdivision. Let us define the necessary notions.

\begin{df}
Let $K$ be an~arbitrary $n$-dimensional geometric simplicial complex.
A~{\bf geometric simplex path} $\Sigma$ is an ordered tuple
$(\sigma_1,\dots,\sigma_l)$ of distinct $n$-simplices of $K$, such
that $\sigma_i\cap\sigma_{i+1}$ is an $(n-1)$-dimensional simplex of
$K$, for all $i=1,\dots,l-1$. 
\end{df}

We say that the path $\Sigma$ has dimension $n$ and length $l$.  The
best way to visualize a geometric simplex path is to imagine
$n$-simplices glued along their boundary $(n-1)$-simplices to form
a~path. Walking along the path from $\sigma_i$ to $\sigma_{i+1}$ may
be visualized as flipping over the $(n-1)$-simplex
$\sigma_i\cap\sigma_{i+1}$. Given a~path
$\Sigma=(\sigma_1,\dots,\sigma_l)$, we define the \emph{interior} of
$\Sigma$ to be the open set
\[\inte\Sigma:=\inte\sigma_1\cup\dots\cup\inte\sigma_l\cup
\inte(\sigma_1\cap\sigma_2)\cup\dots\cup\inte(\sigma_{l-1}\cap\sigma_l).\]

\begin{df}
Assume $\Div(\da^n)$ is a~hereditary chromatic subdivision of the
standard $n$-simplex and $b:V(\Div(\da^n))\ra\{0,1\}$ is a~binary
labeling. A~simplex path $\Sigma=(\sigma_1,\dots,\sigma_l)$ is said to
be in {\bf standard form} if 
\begin{enumerate}
\item[(1)] the number $l$ is even, 
\item[(2)] the end simplices $\sigma_1$ and $\sigma_l$ are 
$0$-monochromatic, 
\item[(3)] the rest of the simplices $\sigma_2,\dots,\sigma_{l-1}$ 
are not monochromatic.
\end{enumerate}
\end{df}

The following result is due to Casta\~neda \& Rajsbaum, \cite{CR2},
where it is formulated, using a~slightly different language, as
a~collection of lemmas.

\begin{thm} \label{thm:cr2-paths}
Assume we are given a~hereditary chromatic subdivision $\Div(\da^n)$,
a~binary labeling $b:V(\Div(\da^n))\ra\{0,1\}$, and an~$n$-dimensional
geometric simplex path $\Sigma$ in standard form. Then there exists
a~chromatic subdivision $\cs(\Div(\da^n))$ of the geometric simplicial
complex $\Div(\da^n)$, such that
\begin{enumerate}
\item[(1)] only the interior of $\Sigma$ is subdivided;
\item[(2)] it is possible to extend the binary labeling $b$ to
  $\cs(\Div(\da^n))$ in such a~way that $\cs(\Sigma)$ has no
  monochromatic $n$-simplices.
\end{enumerate}
\end{thm}
Note, that the obtained subdivision $\cs(\Div(\da^n))$ is
automatically hereditary, since only the simplices in the interior of
the path $\Sigma$, and hence in the interior of the standard simplex
$\da^n$ are subdivided further.

The desire to understand this ``engine'' of the entire proof of
Casta\~neda \& Rajs\-baum has been the driving force behind the
research in this paper. Our purpose here is twofold. On one hand we
want to develop a~self-contained mathematical theory of combinatorial
simplex path subdivisions, motivated by questions in theoretical
distributed computing. On the other hand, we want that our
mathematical theory yields a~simpler, more structural, and more
concise proof of Theorem~\ref{thm:cr2-paths}. The rest of the paper is
devoted to setting up and applying this combinatorial theory.

The specific plan for the paper is as follows. In Chapter 3 we introduce
notations and define the main objects of study: the abstract simplex paths.
In Chapter 4 we define different transformations of these abstract simplex
paths which are of two basic types: the vertex and the edge expansions.
In Chapter 5 more specific transformations are introduced, which we call
summit and plateau moves. This terminology comes from the local shapes
in the height graph of the path. In Chapter 6 we use this toolbox
to prove our main theorem which says that any admissible abstract simplex
path is reducible. We then obtain a new proof of Theorem~\ref{thm:cr2-paths} 
as a~corollary of our combinatorial statement. Finally in Chapter 7
we collect further useful information, which is not directly needed
for the proof of our main result.

An alternative approach to the proof of Casta\~neda and Rajsbaum has
appeared in a~recent work of Attiya, Casta\~neda, Herlihy, and Paz,
see~\cite{ACHP}. An interested reader should consult this paper
as well for a~complete picture.

\section{The combinatorics of path subdivisions} 
\label{sect:3}

In the previous, motivational sections, we have perused the distributed 
computing context and reduced our question to a~completely mathematical
formulation. Within mathematics, all previous work has led one to study subdivisions of 
geometric simplicial complexes, see Theorems~\ref{thm:equiv} and~\ref{thm:cr2-paths}.
In this paper, we make a~further step from geometry to combinatorics.
To this end, we shall define {\it abstract simplex paths}, as well as 
combinatorial operations corresponding to the geometric subdivisions.
We shall then develop combinatorics of path subdivisions
in a~fully self-contained fashion, formally independent on the distributed
computing or geometric context. 

However, before we can proceed with our main definitions, we still need
some notations from standard combinatorics, which we now introduce in 
subsections~\ref{ssect:3.1} and~\ref{ssect:3.2}.

\subsection{Tuples} \label{ssect:3.1} $\,$

\subsubsection{Definition and notations} $\,$

\begin{df}
Let $U$ be an arbitrary set. For an ordered set $S=\{s_1,\dots,s_k\}$, 
with $s_1<\dots<s_k$, an {\bf $S$-tuple} $T$ with elements from $U$ is 
an ordered sequence $T=(a_{s_1},\dots,a_{s_k})$, such that $a_{s_i}\in U$, 
for $1\leq i\leq k$. 
\end{df}
\nin For $s\in S$, we set $T_s:=a_s$. An $S$-tuple
$(a_{s_1},\dots,a_{s_k})$ is said to have {\it length} $k$. We also
write $U(T)$ to denote the universe set associated to an
$S$-tuple~$T$.  We shall now go through further terminology involving
$S$-tuples.

\vskip5pt

\nin {\it Intervals.}  For $i,j\in S$, $i\leq j$, and an $S$-tuple $T$, 
we set $T[i,j]:=(T_i,\dots,T_j)$.

\vskip5pt

\nin {\it $n$-tuples.} For each natural number $n$, we shall for the sake of
brevity say {\it $n$-tuple} instead of $\{1,\dots,n\}$-tuple. Note, that in our 
notations, $n$-tuples and $[n]$-tuples are slightly different mathematical objects.

\vskip5pt

\nin{\it $S$-tuples equal to $0$.} 
We write $T=0$, for an $S$-tuple $T$, if $0\in U(T)$, and $T_s=0$ for
all $s\in S$. 

\vskip5pt

\nin {\it Monochromatic $S$-tuples.}  An~$S$-tuple $T$ is called {\it
  monochromatic} if for some $w\in U(T)$ we have $T_s=w$, for all
$s\in S$. If we need to be more specific we shall say {\it
  $w$-monochromatic}.

\vskip5pt

\nin {\it The height of an $S$-tuple.} 
Assume $T$ is an $S$-tuple with numerical elements, where we include
the case of boolean values by interpreting them as numbers $0$ and
$1$.  We set $h(T):=\sum_{s\in S}T_s$, and call the value $h(T)$ the
{\it height} of~$T$.

\vskip5pt

\nin{\it The index set of occurences of an element in an $S$-tuple.}
Given an $S$-tuple $T$, and an element $q\in U(T)$, we set
\[O(q,T):=(s\in S\,|\,T_s=q).\] 
This is the index set of occurences of $q$ in $T$. We use the round
brackets to emphasize that $O(q,T)$ is an ordered tuple, not just
a~set.

\vskip5pt

\nin{\it Number of occurences of an element in an $S$-tuple.} 
We let $\sharp(q,T)$ denote the number of occurences of $q$
in~$T$, i.e., we set $\sharp(q,T):=|O(q,T)|$.

\vskip5pt

\nin{\it The last occurence of an element in an $S$-tuple.}
We let $\last(q,T)$ denote the index of the last occurence of
  $q$ in $T$, i.e., $\last(q,T):=\max(O(q,T))$. If $O(q,T)=\emptyset$,
  then we set $\last(q,T):=\infty$.

\subsubsection{Concatenation of tuples} $\,$

\vspace{5pt}

\nin
Given two disjoint ordered sets $S'$ and $S''$, we can define a new
ordered set $S=S'\cup S''$, by specifying the new order relation as
follows:
\[s_1<s_2,\textrm{ if }\begin{cases}
s_1,s_2\in S',\textrm{ and } s_1<s_2;\\
s_1,s_2\in S'',\textrm{ and } s_1<s_2;\\
s_1\in S'\textrm{ and }s_2\in S''.
\end{cases}\]
We write $S=S'\circ S''$.  

Assume furthermore, that $T'$ is an $S'$-tuple of length $l'$, and
$T''$ is an $S''$-tuple of length $l''$. We define a new $S$-tuple $T$
by setting $T:=T'\cup T''$ as a~multi-set, and then ordering the
elements according to the order on $S=S'\circ S''$.  We call $T$ the
{\it concatenation} of $T'$ and $T''$ and write $T=T'\circ T''$.

\subsection{The directed edge graph of the cube.} \label{ssect:3.2}

\subsubsection{Definition of $\cuben$} $\,$

\vskip5pt

\nin Assume $n\geq 1$, and let $\cuben$ denote the directed edge graph
of the unit cube in~$\rr^{[n]}$. The vertices of $\cuben$ are labeled
by all $[n]$-tuples of $0$'s and $1$'s, and edges connect those
$[n]$-tuples which differ in exactly one coordinate. We shall view the
$[n]$-tuples of $0$'s and $1$'s as functions $\alpha:[n]\ra[1]$.

$\cuben$ is a~regular graph, where every vertex has both the in-degree
and the out-degree equal to $n+1$. More specifically, for every vertex
$v\in V(\cuben)$, and every $i\in[n]$, there exists a~unique edge
parallel to the $i$th axis, which has $v$ as a~source. In total, the
graph $\cuben$ has $2^{n+1}$ vertices and $(n+1)\cdot 2^{n+1}$ edges.

We shall use the standard graph terminology applied to the graph
$\cuben$ such as {\it directed cycle} and {\it directed path}. In
particular, a~{\it partial matching} on $\cuben$ is a triple
$(A,B,\varphi)$, where $A$ and $B$ are disjoint sets of vertices of
$\cuben$, and $\varphi:A\ra B$ is a bijection, such that for each
$v\in A$, the vertices $v$ and $\varphi(v)$ are connected by an edge
(in this case they are of course automatically connected by~$2$
edges).

\subsubsection{Cube loops}

\begin{df}\label{df:loop}
An {\bf $[n]$-cube loop} is a~directed cycle in $\cuben$, which
contains the origin, and does not have any self-intersections.
\end{df}

Clearly, to describe an $[n]$-cube loop one can start at the origin
and then list the directions of the edges as we trace the loop. This
procedure will yield a~$t$-tuple $(q_1,\dots,q_t)$, with $q_i\in[n]$,
for all $1\leq i\leq t$. Conversely, given such a~$t$-tuple
$Q=(q_1,\dots,q_t)$, it describes an $[n]$-cube loop if and only if
the following conditions are satisfied:
\begin{enumerate}
\item[(1)] {\it the cycle condition:} for every $l\in[n]$, the number
  $\sharp(l,Q)$ is even;
\item[(2)] {\it no self-intersections:} for all $1\leq i<j\leq t$,
  such that $(i,j)\neq (1,t)$, there exists $l\in[n]$, such that the
  number $\sharp(l,Q[i,j])$ is odd.
\end{enumerate}

\begin{figure}[hbt]
\begin{center}

  \input{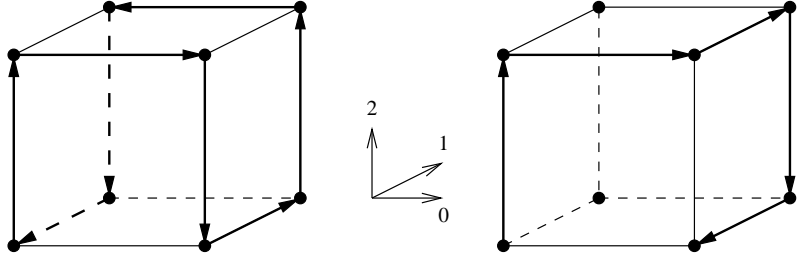}  

\end{center}
\caption{A $[2]$-cube loop, and a $[2]$-cube $0$-path.}
\label{fig:cube}
\end{figure}

In the continuation, we shall usually describe an~$[n]$-cube loop by
taking a~tuple $(q_1,\dots,q_t)$ satisfying the conditions above.  An
example of a~$[2]$-cube loop is shown on the left-hand side of
Figure~\ref{fig:cube}. The corresponding $t$-tuple is
$(2,0,2,1,2,0,2,1)$.

\subsubsection{Cube paths}

\begin{df}\label{df:path}
Assume $p\in[n]$. An {\bf $[n]$-cube $p$-path} is a~directed path $I$
in the graph $\cuben$, such that
\begin{itemize} 
\item $I$ starts at the origin, and ends in the vertex
  $(0,\dots,0,1,0,\dots,0)$, where $1$ is in position~$p$;
\item $I$ does not have any self-intersections;
\item $I$ contains exactly one edge parallel to the $p$th axis.
\end{itemize}
\end{df}

In line with the situation of the $[n]$-cube loops, the $[n]$-cube
$p$-path is given by a~tuple $(q_1,\dots,q_t)$, where 
$q_1,\dots,q_t\in[n]\setminus\{p\}$, together with a~number
$1\leq s\leq t-1$, subject to the following conditions:
\begin{enumerate}
\item[(1)] {\it the path condition:} for every $l\in[n]\sm\{p\}$, the
  number $\sharp(l,Q)$ is even;
\item[(2)] {\it no self-intersections:} for all $1\leq i<j\leq s$,
  there exists $l\in[n]\setminus\{p\}$, such that the number
  $\sharp(l,Q[i,j])$ is odd; the same is true for all $s+1\leq i<j\leq
  t$.
\end{enumerate}

An example of a~$[2]$-cube $0$-path is shown on the right-hand
side of Figure~\ref{fig:cube}. The corresponding $t$-tuple is
$(2,1,2,1)$, and $s=1$.

\subsection{The abstract simplex paths} $\,$

\subsubsection{Definition and associated data.} $\,$

\vspace{5pt}

\noindent
Abstract simplex paths are the main objects of study in this paper.

\begin{df}
Let $k$ and $n$ be positive integers.  An {\bf abstract simplex path}
$P$ is a triple $(I,C,V)$, where $I$ is an $[n]$-tuple of boolean
values, $C$ is a~$(k-1)$-tuple of elements of $[n]$, and $V$ is
a~$(k-1)$-tuple of boolean values. We request that the $(k-1)$-tuple
$C$ satisfies an additional ``no back-flip'' condition: for all
$i=1,\dots,k-2$, we have $C_i\neq C_{i+1}$.
\end{df} 

Given an abstract simplex path $P=(I,C,V)$ as above, the number $k$ is
called its {\it length}, the number $n$ is called its {\it dimension},
and the tuple $I$ is called {\it the initial simplex} of $P$. When we
need to connect back to the path name we shall also write $I(P)$,
$C(P)$, and $V(P)$.

Since the set $[n]$ indexes the vertices of an $n$-simplex, and the
set $[1]$ indexes boolean values, each pair $(C_i,V_i)$ can be thought
of as assigning a chosen boolean value $V_i$ to a~chosen vertex $C_i$
of an $n$-simplex. So, the path $P$ can be interpreted as starting
from the initial $[n]$-tuple $I(P)$, and then changing specific
boolean entries according to the pattern given by $C(P)$ and $V(P)$.
We call a~path $P$ a~$0$-path if $I(P)=0$ and $V(P)=0$.


Next, we proceed to describe how to associate certain {\it data} to
every abstract simplex path.

\begin{df} 
Assume $P=(I,C,V)$ is a path of dimension $n$ and length~$k$. Fix
$1\leq j\leq k$. For every $i\in[n]$, we
set \[e_i^j(P):=V_{\last(i,C(1,j-1))}.\] If
$\last(i,C(1,j-1))=\infty$, we set $e_i^j(P):=I_i$. We then set
\[R^j(P):=(e_0^j(P),\dots,e_n^j(P)).\] 
We call the $[n]$-tuples $R^1(P)$, $\dots$, $R^k(P)$ the {\bf
  simplices of the path}~$P$.  Finally, we set
\[R(P):=(R^1(P),\dots,R^k(P)).\]
\end{df}
 Note that for $j=1$, the set $\{1,\dots,j-1\}$ is empty, and
 $\last(i,C(1,0))=\infty$, for all $i$, implying that $R^1(P):=I(P)$.

\subsubsection{Concatenation of paths}
\begin{df}\label{df:conc}
Assume that $P$ and $Q$ are paths of dimension $n$ of lengths $k$
and~$l$. A path $T$ of dimension $n$ is called a~{\bf concatenation}
of $P$ and $Q$ if
\begin{itemize}
\item $I(T)=I(P)$, $R^k(T)=R^{k+1}(T)=I(Q)$, 
\item $C(T)=C(P)\circ(p)\circ C(Q)$, for some
$p\in[n]$, 
\item $V(T)=V(P)\circ(I(Q)_p)\circ V(Q)$.
\end{itemize}
\end{df}
Note that the fact that $P$ in Definition~\ref{df:conc} is
a~well-defined path implies that $p\neq C(P)_{k-1}$ and $p\neq
C(Q)_1$. Since several conditions need to be satisfied, it might very
well happen that there are no concatenations of $P$ and $Q$. On the
other hand, there could be several choices of $p$, for which all the
conditions of Definition~\ref{df:conc} are satisfied, so there may
exist several concatenations of $P$ and $Q$. Iterating
Definition~\ref{df:conc} we can talk about concatenations of any
finite number of paths.

\subsubsection{The actions of symmetric groups on the sets of paths.} $\,$

\vspace{5pt}

\nin
There is a natural group action of the symmetric group $\cs_{[n]}$ of
all permutations of the set $[n]$ on the set of all paths of dimension
$n$: namely, for a~permutation $\pi\in\cs_{[n]}$, viewed as
a~bijection $[n]\ra[n]$, and a~path $P=(I,C,V)$ of dimension $n$ and
length $k$, we set $\pi(P):=(\pi(I),\pi(C),V)$, where
\[\pi(I)=(I_{\pi(0)},\dots,I_{\pi(n)}),\]
and
\[\pi(C):=(\pi(C_1),\dots,\pi(C_{k-1})).\]

We may also consider the reflection action of $\cs_2$ on the paths
which switches the direction, which is easiest to describe in the
$R(P)$-notations:
\[(R^1(P),\dots,R^k(P))\mapsto(R^k(P),\dots,R^1(P)).\]
Alternatively, the nontrivial element of $\cs_2$ takes the path
$P=(I,C,V)$ to the path $\bar P=(\bar I,\bar C,\bar V)$, where $\bar
I=R^k(P)$, 
\[\bar C=(C_{k-1},\dots,C_1),\] and
\[\bar V=(e_{C_{k-1}}^{k-1},\dots,e_{C_1}^1).\]

\subsubsection{Atomic and admissible paths.} $\,$

\vspace{5pt}

\nin
In this paper we are concerned with algorithms transforming certain
types of paths, which we now proceed to define.

\begin{df}
Let $P$ be an~abstract simplex path $P$ of dimension $n$ and length $k$.
\begin{enumerate} 
\item[(1)] The path $P$ is called {\bf atomic} if
\begin{itemize}
\item $k$ is even; 
\item we have $I(P)=R^k(P)=0$, and these are the only monochromatic
  simplices of $P$.
\end{itemize}
\item[(2)] The path $P$ is called {\bf admissible} if it is a
  concatenation of atomic paths.
\end{enumerate}
\end{df}

Since for atomic and admissible paths $P$ we have $I(P)=0$, we shall
at times for brevity skip it from the definition and just write
$P=(C,V)$.  Given an admissible path $P$, the set of atomic paths,
whose concatenation is $P$, is uniquely defined. Furthermore, the
actions of $\cs_{[n]}$ and $\cs_2$ from above restrict to the subsets
of atomic and admissible paths.

It is often useful to complement the string notation of the path with
a more graphic representation, by listing the parts of $[n]$-tuples
$R^1(P)$, $\dots$, $R^k(P)$ where the values are changed at some point
along the path, and indicating, using the downward arrows, the
positions where the boolean values are changed. Many examples of
atomic abstract simplex paths given in that form can be found in the
Table~\ref{table:p6d}.

\subsubsection{Connection between geometric simplex paths with boolean labels 
and abstract simplex paths.} \label{sssect:3.3.5} $\,$

\vskip5pt

\nin To any given geometric simplex path with boolean labels
$\Sigma=(\sigma_1,\dots,\sigma_l)$ one can associate a~unique abstract
simplex path $P=(I,C,V)$. To do that, simply take the $[n]$-tuples of
labels of the simplices $\sigma_1,\dots,\sigma_l$ to be the tuples
$R^1(P),\dots,R^l(P)$. Alternatively, we an take $I(P)$ to be the
labels of $\sigma_1$, and then record in $C(P)$ the indices (colors)
of the vertices which are flipped as we walk along the path, and
record in $V(P)$ the labels assigned to the new vertices after each
flip. Note, that under this correspondence, the geometric simplex
paths in standard form will yield an atomic abstract simplex path.

\begin{figure}[hbt]
\begin{center}

  \input{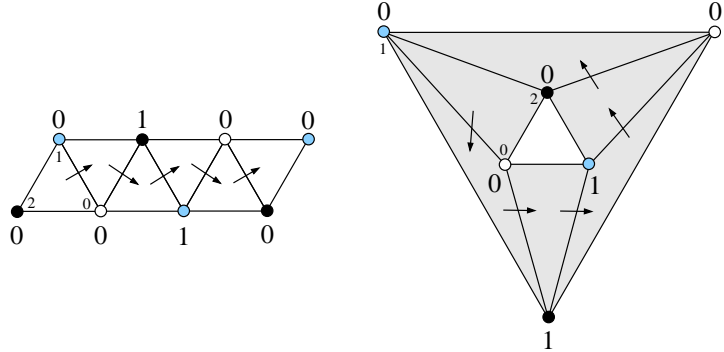}  

\end{center}
\caption{Two geometric simplex paths with boolean labeling corresponding to
the abstract simplex path on Figure~\ref{tab:ex335}.}
\label{fig:ex335}
\end{figure}

\begin{figure}[hbt]
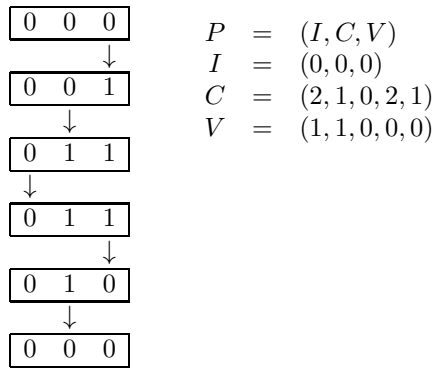

\begin{minipage}{3cm}
\[
\begin{array} {ccc}
\three{0}{0}{0}
 \darc  \\
\three{0}{0}{1}
 \darb  \\
\three{0}{1}{1}
\dara \\
\three{0}{1}{1}
 \darc  \\
\three{0}{1}{0}
 \darb \\
\three{0}{0}{0}
\end{array}
\]
\end{minipage}
\begin{minipage}{3cm}
\[\begin{array}{ccl}
P&=&(I,C,V) \\ 
I&=&(0,0,0)  \\
C&=&(2,1,0,2,1)  \\
V&=&(1,1,0,0,0) \\ \\ \\  \\ \\ \\ \\ \\
\end{array}\]
\end{minipage}
\caption{The abstract simplex path corresponding to the geometric simplex
paths with boolean labelings on Figure~\ref{fig:ex335}.}
\label{tab:ex335}
\end{figure}

Reversely, given an abstract simplex path, one can always associate to
it the geometric simplex path with boolean labels, though this time
the geometric path will not be unique. Figure~\ref{fig:ex335} shows
two different geometric simplex paths with boolean labels, which have
the same associated abstract simplex path, described by
Figure~\ref{tab:ex335}.

\subsubsection{The height data.} $\,$

\vspace{5pt}

\begin{figure}[hbt]
\begin{center}

  \input{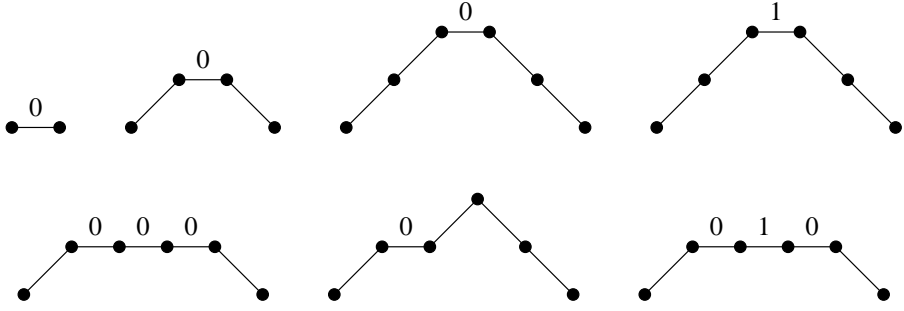}  

\end{center}
\caption{The height graphs.}
\label{fig:gr1}
\end{figure}

\nin Further data, which we associate to an abstract simplex path $P$,
is its {\it height graph} $h(P)$.  Specifically, for a~path $P$ of
length $k$, we set $h_i(P):=h(R^i(P))$, for $i=1,\dots,k$. When $P$ is
atomic, we have $h_1(P)=h_k(P)=0$, and $h_i(P)>0$ for $1<i<k$. The
height graph of $P$ is a~partially marked graph imbedded in $\rr^2$,
whose set of vertices is $\{(i,h_i(P))\,|\,i=1,\dots,k\}$. The set of
edges of $h(P)$ is described as follows: for each $i=1,\dots,k-1$ we
connect the dots $(i,h_i(P))$ and $(i+1,h_{i+1}(P))$ by an interval,
and in addition, if $h_i(P)=h_{i+1}(P)$, i.e., if $V_i=e_{C_i}^i$, we
mark the corresponding edge with $V_i$. The examples of height graphs
of atomic paths of length $2$, $4$, and $6$ are shown on
Figure~\ref{fig:gr1}; e.g., the height graph of the abstract simplex path
in Figure~\ref{tab:ex335} is the third graph in Figure~\ref{fig:gr1}. 

For future reference, we introduce the height statistics
$H(P):=(\nu_0,\dots,\nu_{n+1})$, where
$\nu_j=\sharp(j,(h_1(P),\dots,h_k(P)))$. Since $P$ is atomic, we have
$\nu_0=2$, and $\nu_{n+1}=0$. We order such height statistics using
what is commonly known as the reverse lexicographic order on
$(n+2)$-tuples, i.e.,
$(\nu_0,\dots,\nu_{n+1})\succ(\mu_0,\dots,\mu_{n+1})$ if and only if
there exists $0\leq i\leq n+1$, such that $\nu_i>\mu_i$, and
$\nu_{i+1}=\mu_{i+1}$, $\dots$, $\nu_{n+1}=\mu_{n+1}$.


\section{Combinatorial deformations of an abstract simplex path}

\noindent
At this point, we have introduced the abstract simplex paths and the
associated data.  We have seen different presentations for that data
and we classified this information for all atomic paths of length at
most~$6$. Next, we describe two types of path deformations, which we
call {\it path expansions}. The two specific types which we have in
mind are called {\it vertex expansion} and {\it edge expansion}.


\subsection{Vertex expansion.} 

\subsubsection{Combinatorial definition of the vertex expansion.} $\,$

\vspace{5pt}

\noindent
Assume $P=(I,C,V)$ is an~abstract simplex path of length $k\geq 3$ and
dimension~$n$. The input data for the vertex expansion consists of:
\begin{itemize} 
\item a number $m$, such that $2\leq m\leq k-1$;
\item an~$[n]$-tuple $D=(d_0,\dots,d_n)$ of boolean values;
\item an~$[n]$-cube loop $Q=(q_1,\dots,q_t)$, such that $q_1=C_{m-1}$
and $q_t=C_m$.
\end{itemize}

The number $m$ is the position of the expansion. For convenience of
notations, we set $(c_0,\dots,c_n):=R^m(P)$.

\begin{df}\label{df:vexp}
The vertex expansion of a~path $P=(I,C,V)$ with respect to the input
data $(m,D,Q)$ is a new path $\wti P$ of dimension $n$ defined as
follows:
\begin{itemize}
\item $I(\wti P)=I$;
\item $C(\wti P)$ is obtained from $C$ be inserting the tuple
$(q_2,\dots,q_{t-1})$ between $C_{m-1}$ and $C_m$; 
\item $V(\wti P)$ is obtained from $V$ by replacing $V_{m-1}$ with the tuple
 $w_1,\dots,w_{t-1}$, where 
\begin{equation}\label{eq:wi1}
w_i=\begin{cases}
c_{q_i}, & \textrm{ if }\,\,\, \sharp(q_i,(q_1,\dots,q_i)) \textrm{ is even,}\\
d_{q_i}, & \textrm{ if }\,\,\, \sharp(q_i,(q_1,\dots,q_i)) \textrm{ is odd,}
\end{cases}
\end{equation}
for all $i=1,\dots,t-1$.
\end{itemize}
\end{df}
We shall write $P(m,D,Q)$ to denote the obtained path $\wti P$.
Clearly, path $P(m,D,Q)$ has length $k+t-2$.

\subsubsection{Geometric interpretation of the vertex expansion.} 
\label{ssect:412} $\,$

\vskip5pt

\nin Let us assume we have an abstract simplex path $P=(I,C,V)$.  We
want to see what a~vertex expansion $P(m,D,Q)$ corresponds to
geometrically. To do that, we need a~specific subdivision of the
interior of the simplex $\sigma_m$. There are several equivalent ways
to view this subdivision.  One could use so-called chromatic joins,
see~\cite{CR2,HKR}.  Alternatively, one can use Schlegel diagrams, as
was done in \cite{subd}, see also \cite{Coxeter,Grun} for the polytope
background. We prefer the latter, as then the fact that we actually
have a subdivision is immediate. In this language, we simply replace
the simplex $\sigma_m$ by the Schlegel diagram $S$ of the
$(n+1)$-dimensional cross-polytope, see \cite{subd} for details. In
this subdivision there are $n+1$ new vertices, which are in natural
bijection with the vertices of $\sigma_m$: each vertex $v$ of
$\sigma_m$ has a unique opposite vertex $\op(v)$ in the corresponding
cross-polytope. The vertex of $S$ which is opposite to $v\in
V(\sigma_m)$ gets the same color as $v$, which immediately implies
that the obtained subdivision is chromatic. The $[n]$-tuple $D$
encodes the extension of the binary labeling to $S$ as follows: let
$v\in V(\sigma_m)$ have the color $\rho$, then the label of $\op(v)$
is~$d_\rho$.

This describes the new subdivision of $\sigma_m$. The remaining
variable $Q$ describes how the new path $\widetilde\Sigma$ goes
through this subdivision. In the old path, one simply flipped from
$\sigma_{m-1}$ to $\sigma_m$, and then on to~$\sigma_{m+1}$. In the
new subdivision, one has to connect $\sigma_{m-1}$ to $\sigma_{m+1}$
by flipping through~$S$. To do that, we need to understand the
combinatorics of~$S$ better. The standard fact about the
$(n+1)$-dimensional cross-polytope is that it is dual to the
$[n]$-cube, see~\cite{Coxeter,Grun}. One implication of that duality
is that $n$-simplices of the cross-polytope correspond to the vertices
of the $[n]$-cube, and a~geometric $n$-simplex path on the boundary of
the cross-polytope corresponds to a~regular edge path on the boundary
of an~$[n]$-cube. The new path $\widetilde\Sigma$ is supposed to enter
the Schlegel diagram $S$ from one side, and then traverse it and exit
on one of the other sides. Clearly, combinatorially we can think of it
as an~$n$-simplex loop which starts and ends at the $n$-simplex $F$ of
the cross-polytope, which is taken as the basis for the Schlegel
diagram. Dually, this corresponds to an~$[n]$-cube loop,
cf.\ Definition~\ref{df:loop}, which is precisely the loop~$Q$.

Finally, the identity~\eqref{eq:wi1} describes the change of the
labels correctly, as the parity of the number
$\sharp(q_i,(q_1,\dots,q_i))$ tells us whether we get the vertex in
the base face of the Schlegel diagram, this happens if the number is
even, or we get one of the opposite vertices, lying in the interior of
the subdivided simplex, which happens if the number is odd.

\subsubsection{An example.} \label{ssect:413} $\,$

\vskip5pt

\begin{figure}[hbt]
\begin{center}

  \input{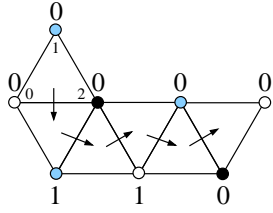}  

\end{center}
\caption{A geometric simplex path corresponding to the abstract
  simplex path with $I=(0,0,0)$, $C=(1,0,1,2,0)$, $V=(1,1,0,0,0)$.}
\label{fig:exve1}
\end{figure}

\nin Let us illustrate this interpretation by an example. Consider an
abstract simplex path $P=(I,C,V)$, with $I=(0,0,0)$, $C=(1,0,1,2,0)$,
and $V=(1,1,0,0,0)$. One of the corresponding geometric simplex paths
is shown on Figure~\ref{fig:exve1}.

Consider the~vertex expansion $P(m,D,Q)$, for $m=3$, $D=(0,0,0)$,
$Q=(0,1,2,0,2,1)$. By Definition~\ref{df:vexp} we have
$P(m,D,Q)=(I,\widetilde C,\widetilde V)$, and we now proceed to determine
$\widetilde C$ and $\widetilde V$. We have the following data which we 
substitute in this definition: $C_{m-1}=C_2$, $C_m=C_3$, and $t=6$.
The rule in Definition~\ref{df:vexp} says that to obtain $\widetilde C$
we need to insert $(q_2,\dots,q_5)=(1,2,0,2)$ between $C_2$ and $C_3$ 
in~$C$. This gives us the answer $C=(1,0,1,2,0,2,1,2,0)$. Furthermore, 
to get $\widetilde V$ from $V$, we need to replace $V_2=1$ with the tuple 
$(w_1,\dots,w_5)$. To determine that tuple, consider the table~\eqref{eq:a1}.
\begin{equation}\label{eq:a1}
\begin{array} {|c||c|c|c|c|c|}
\hline
i   & 1 & 2 & 3 & 4 & 5 \\ \hline
q_i & 0 & 1 & 2 & 0 & 2 \\ \hline
w_i & d_0 & d_1 & d_2 & c_0 & c_2 \\ \hline
\end{array}
\end{equation}

\begin{figure}[hbt]
\begin{center}

  \input{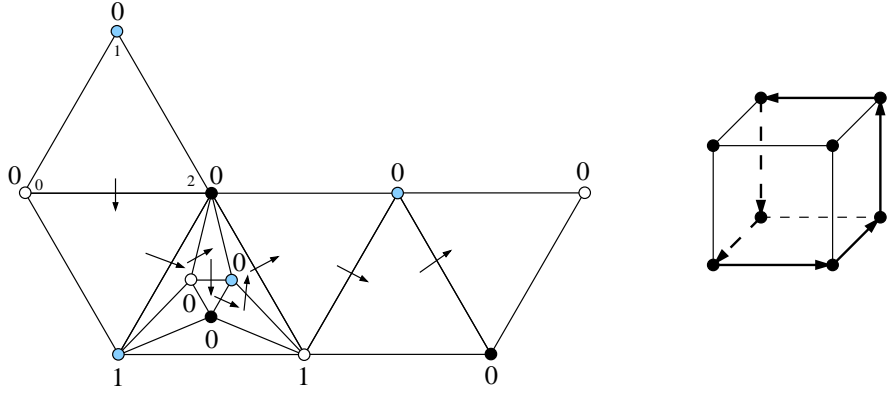}  

\end{center}
\caption{A vertex expansion of the geometric simplex path from 
Figure~\ref{fig:exve1}, and the corresponding $[2]$-cube loop.}
\label{fig:exve2}
\end{figure}

The second row in that table simply shows the relevant part of $Q$. 
The third row shows which value is assigned to $w_i$ by 
the rule~\eqref{eq:wi1}. Since $(c_0,c_1,c_2)=R^2(P)=(1,1,0)$ and 
$(d_0,d_1,d_2)=(0,0,0)$, we arrive at $(w_1,\dots,w_5)=(0,0,0,1,0)$, 
which then yields the answer $\widetilde V=(1,0,0,0,1,0,0,0,0)$. 
An associated geometric simplex path is shown on Figure~\ref{fig:exve2}.

\subsection{Edge expansion.} 

\subsubsection{Combinatorial definition of the edge expansion.} $\,$

\vskip5pt

\noindent
Assume again that $P=(I,C,V)$ is an abstract simplex path of length
$k\geq 4$ and dimension~$n$. The input data for the edge expansion
consists of:
\begin{itemize} 
\item a number $m$, such that $2\leq m\leq k-2$;
\item an~$([n]\setminus\{C_m\})$-tuple $D=(d_0,\dots,\widehat d_{C_m},\dots,d_n)$ 
of boolean values;
\item an~$[n]$-cube $C_m$-path given by $Q=(q_1,\dots,q_t)$, such that $q_1=C_{m-1}$
and $q_t=C_{m+1}$, together with a~number $1\leq s\leq t-1$.
\end{itemize}

For convenience, the alternative notation
$D=(d_0,\dots,d_{C_m-1},-,d_{C_m+1},\dots,d_n)$ will also be used.
Again, we set $(c_0,\dots,c_n):=R^m(P)$.

\begin{df}\label{df:eexp}
The edge expansion of a~path $P$ with respect to the input data
$(m,D,Q,s)$ is a new path $\wti P$  of dimension $n$ defined as
follows:
\begin{itemize}
\item $I(\wti P)=I$;
\item $C(\wti P)$ is obtained from $C$ be inserting the tuple
  $(q_2,\dots,q_s)$ between $C_{m-1}$ and $C_m$, and then inserting
  the tuple $(q_{s+1},\dots,q_{t-1})$ between $C_m$ and $C_{m+1}$;
\item $V(\wti P)$ is obtained from $V$ by replacing $V_{m-1}$ with the
  tuple $w_1,\dots,w_s$, and then inserting the tuple
  $w_{s+1},\dots,w_{t-1}$ between $V_m$ and $V_{m+1}$, where
\begin{equation}\label{eq:wi2}
w_i=\begin{cases}
c_{q_i}, & \textrm{ if }\,\,\, \sharp(q_i,(q_1,\dots,q_i)) \textrm{ is even,}\\
d_{q_i}, & \textrm{ if }\,\,\, \sharp(q_i,(q_1,\dots,q_i)) \textrm{ is odd,}
\end{cases}
\end{equation}
for all $i=1,\dots,t-1$.
\end{itemize}
\end{df}

We shall write $P(m,D,Q,s)$ to denote the obtained path $\wti P$.
Clearly, the path $P(m,D,Q,s)$ has length $k+t-2$.

\subsubsection{Geometric interpretation of the edge expansion.} 
\label{ssect:422} $\,$

\vskip5pt

\nin Let us describe the geometry behind the edge expansions. What
happens in this case is that we first subdivide the $(n-1)$-simplex
$\sigma_m\cap\sigma_{m+1}$, call this subdivision~$S$. Then, we extend
this to a~subdivision of $\sigma_m$ by coning over $S$ with the apex
being the vertex $v_m=\sigma_m\setminus(\sigma_m\cap\sigma_{m+1})$,
and we extend it to a~subdivision of $\sigma_{m+1}$ by coning over $S$ 
with the apex at $v_{m+1}=\sigma_{m+1}\setminus(\sigma_m\cap\sigma_{m+1})$. 
In total, only the interiors of the simplices $\sigma_m$, $\sigma_{m+1}$,
and $\sigma_m\cap\sigma_{m+1}$ get subdivided, and one may view the whole
thing as a~suspension of~$S$. The subdivision $S$ itself is again
chosen to be a~Schlegel diagram of a~cross-polytope of appropriate
dimension: here we are dealing with the cross-polytope of dimension~$n$. 
In this situation, $C_m$ is the color of the vertex $v_m$, which by the way is 
the same as the color of the vertex $v_{m+1}$. The $[n]\setminus\{C_m\}$-tuple
$D$ gives the labels of the new vertices of the subdivision~$S$.

In analogy to the analysis of the vertex expansion, we see that the
$(n-1)$-simplices of $S$ are in one-to-one correspondance with the
vertices of the $[n]\setminus\{C_m\}$-cube. The $n$-simplices of the
new subdivision of $\sigma_m\cup\sigma_{m+1}$ are obtained from the
$(n-1)$-simplices of the Schlegel diagram $S$ by adding either $v_m$
or $v_{m+1}$, so there are exactly $2\cdot(2^n-1)$ of them. Thus, the
new path $\widetilde\Sigma$, which is described by the edge expansion,
will proceed as follows:
\begin{itemize}
\item it will enter the subdivision of $\sigma_m$, and flip for
  a~while among the $(n-1)$-simplices of $S$, keeping $v_m$ as one of
  the vertices;
\item it will flip over to the subdivision of $\sigma_{m+1}$, swapping
  $v_m$ for $v_{m+1}$;
\item finally, it will flip for a~while among the $(n-1)$-simplices of
  $S$, keeping $v_{m+1}$ as one of the vertices, and then exit at the
  same $(n-1)$-simplex of $\sigma_{m+1}$ as $\Sigma$ did.
\end{itemize}
This data is described by the variables $Q$ and $s$. The part
$(q_1,\dots,q_s)$ of $Q$ describes an~$(n-1)$-simplex path in $S$
corresponding to the $n$-simplex path in the total subdivision, with
all simplices having $v_m$ as a~vertex. 

\subsubsection{An example.} \label{ssect:423} $\,$

\vskip5pt

\begin{figure}[hbt]
\begin{center}

  \input{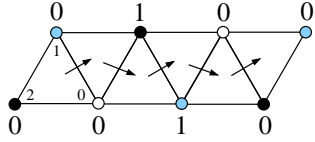}  

\end{center}
\caption{A geometric simplex path corresponding to the abstract
  simplex path with $I=(0,0,0)$, $C=(2,1,0,2,1)$, $V=(1,1,0,0,0)$.}
\label{fig:exee1}
\end{figure}

\nin Again, we would like to illustrate how the edge expansion works
by an example. Consider an abstract simplex path $P=(I,C,V)$, with
$I=(0,0,0)$, $C=(2,1,0,2,1)$, and $V=(1,1,0,0,0)$. One of the
corresponding geometric simplex paths is shown on
Figure~\ref{fig:exee1}.

Consider the~edge expansion $P(m,D,Q,s)$, where $m=3$, $D=(-,0,1)$,
$Q=(1,2,1,2)$, and $s=2$. By Definition~\ref{df:eexp}, 
we have $P(m,D,Q,s)=(I,\widetilde C,\widetilde V)$, and we now determine 
the tuples $\widetilde C$ and~$\widetilde V$. Here we have $q_s=q_2$, 
$C_m=C_3$, $t=4$, so $q_{s+1}=q_{t-1}=q_3$.
By Definition~\ref{df:eexp}, to get $\widetilde C$ from $C$ we need
to insert $q_2$ between $C_2$ and $C_3$, and insert $q_3$ between 
$C_3$ and $C_4$. As a result we get $\widetilde C=(2,1,2,0,1,2,1)$.
To get $\widetilde V$ from $V$, we need to replace $V_2$ with $(w_1,w_2)$,
and insert $w_3$ between $V_3$ and $V_4$. To calculate the tuple
$(w_1,w_2,w_3)$ consider the table~\eqref{eq:a2}.
\begin{equation}\label{eq:a2}
\begin{array} {|c||c|c|c|}
\hline
i   & 1 & 2 & 3 \\ \hline
q_i & 1 & 2 & 1 \\ \hline
w_i & d_1 & d_2 & c_1 \\ \hline
\end{array}
\end{equation}

\begin{figure}[hbt]
\begin{center}

  \input{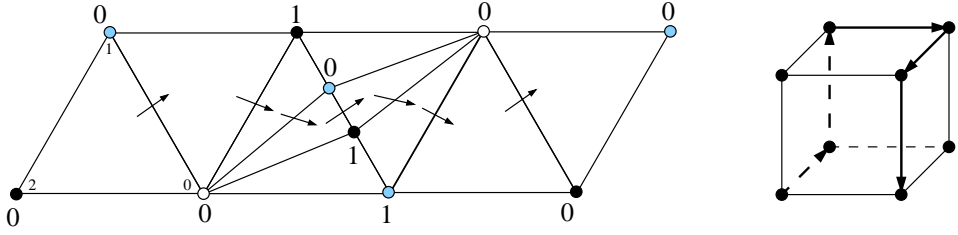}  

\end{center}
\caption{An edge expansion of the geometric simplex path from 
Figure~\ref{fig:exee1}, and the corresponding $[2]$-cube loop.}
\label{fig:exee2}
\end{figure}

The second row in that table again shows the relevant part of $Q$. 
The third row shows which value is assigned to $w_i$ by 
the rule~\eqref{eq:wi2}. Since $(c_0,c_1,c_2)=R^3(P)=(1,1,0)$ and 
$(d_0,d_1,d_2)=(-,0,1)$, we arrive at $(w_1,w_2,w_3)=(0,1,1)$, 
which then yields the answer $\widetilde V=(1,0,1,0,1,0,0)$. The
resulting geometric simplex path is shown on Figure~\ref{fig:exee2}.

\subsection{The exhaustive expansion.}

\subsubsection{Combinatorial definition of the exhaustive expansion.}$\,$

\vskip5pt

\nin Let $P$ be an abstract simplex path, and let $\wti P$ be either
a~vertex expansion $P(m,D,Q)$ or an edge expansion $P(m,D,Q,s)$. We
describe how to associate a~$[1]\times[n]    $-array $A$ to that pair
$(P,\wti P)$. To start with, in any case we set
$A[0,-]:=R^m(P)$. Proceeding to the second row: for a vertex expansion
$P(m,D,Q)$ we set $A[1,-]:=D$, while for an edge expansion
$P(m,D,Q,s)$ we set $A[1,i]:=D_i$, for all $i\in[n]\setminus\{C_m\}$,
and $A[1,C_m]=V_m$.

Since the vertices of $\cuben$ are indexed by all functions
$\alpha:[n]\ra[1]$, the $[1]\times[n]$-array of boolean values $A$ can
be used to associate boolean $[n]$-tuples to vertices of
$\cuben$. Specifically, to a~vertex $\alpha:[n]\ra[1]$ of $\cuben$ we
associate the $[n]$-tuple
$A(\alpha):=(A[\alpha(0),0],A[\alpha(1),1],\dots,A[\alpha(n),n])$. 

Let $M(A)$ denote the set of all $\alpha:[n]\ra[1]$, such that
$A(\alpha)$ is monochromatic. If $M(A)\neq\emptyset$ there are two possibilities. The first option is that $M(A)$ consists of two 
``opposite'' $[n]$-tuples, one is $0$-monochromatic, and the other 
one is $1$-monochromatic. The second option is that all
$\alpha\in M(A)$ are $b$-monochromatic, where $b\in[1]$. In this 
case, the subgraph of $\cuben$ induced by $M(A)$ is isomorphic to 
$\textrm{Cube}_{[m]}$, for some $m\leq n$; in particular, $|M(A)|$ 
is a~power of~$2$.

Intuitively, we shall call the expansion $\wti P$ {\it exhaustive} if
all vertices $\alpha\in M(A)$ either belong to $\wti P$, or can be
matched (in the sense of graph theory) to each other. Unfortunately,
it is somewhat technical to express the fact that a~vertex $\alpha$ 
belongs to the new path $\wti P$. We now give the formal definition.
 
\begin{df} \label{df:exh} 
Let the paths $P$ and $\wti P$ be as above. The expansion $\wti P$ is
called {\bf exhaustive} if, there exists a~partial matching on the
vertices in $M(A)$, such that, for each unmatched $\alpha\in M(A)$,
there exists $i$ satisfying the following conditions:
\begin{itemize} 
\item if $\wti P$ is a vertex expansion, then $1\leq i\leq t-1$;
\item if $\wti P$ is an edge expansion and $\alpha(C_m)=0$, then 
$1\leq i\leq s$;
\item if $\wti P$ is an edge expansion and $\alpha(C_m)=1$, then 
$s+1\leq i\leq t$;
\item for all $j\in[n]$, we have 
\[\alpha(j)=\sharp(j,(q_1,\dots,q_i))\mod 2.\]
\end{itemize}
\end{df}

\subsubsection{Geometric interpretation of exhaustive vertex expansion.}$\,$

\vskip5pt

\nin Let us understand what it means geometrically for the vertex expansion 
to be exhaustive. For this, recall the framework of the 
subsubsection~\ref{ssect:412}, where the vertex expansion was interpreted 
as a~replacement of an $n$-simplex $\sigma_m$ by a~Schlegel diagram $S$ 
of an~$(n+1)$-dimensional cross-polytope. The $n$-simplex $\sigma_m$ 
corresponded to the face $F$ of the cross-polytope, with regard to which 
the Schlegel diagram was taken.
 
In this case, the $[1]\times[n]$-array $A$ contains all
the labels of the vertices of that cross-polytope. All the functions
$\alpha:[n]\ra[1]$ correspond to $n$-faces of the cross-polytope, and,
accordingly, the $[n]$-tuples
$A(\alpha)=(A[\alpha(0),0],\dots,A[\alpha(n),n])$ correspond to labels
of these faces. There are $2^{n+1}$ of these $n$-faces, and all but
$F$, which corresponds to $\alpha\equiv 0$, are $n$-simplices in the
Schlegel diagram~$S$. Recall, that $\sigma_m$ itself is not
monochromatic, so the set $M(A)$ corresponds to all monochromatic
simplices of~$S$. Definition~\ref{df:exh} then says precisely that all
the monochromatic simplices in $S$, either belong to the new
path, or can be matched pairwise, so that each two
matched $n$-simplices share an~$(n-1)$-dimensional boundary simplex.

\subsubsection{Geometric interpretation of exhaustive edge expansion.}$\,$

\vskip5pt

\nin Here we are in the framework discussed in the 
subsubsection~\ref{ssect:422}. We have two $n$-simplices $\sigma_m$ 
and $\sigma_{m+1}$ on our path. First the $(n-1)$-simplex 
$\sigma_m\cap\sigma_{m+1}$ is replaced by a~Schlegel diagram, then 
this subdivision is extended to the interior of the simplices 
$\sigma_m$ and $\sigma_{m+1}$. As in the case of vertex expansion, 
the edge expansion is exhaustive, if all monochromatic simplices 
in the new subdivision of $\sigma_m\cup\sigma_{m+1}$ are either 
on the new path, or can be matched to each other, having a~common
$(n-1)$-simplex in each matched pair.

\begin{example}\label{ex:p4}
{\it An example of an exhaustive edge expansion.} As an example let us
consider an atomic path P of dimension $n\geq 2$ and length $4$, given
by $P=((0,1,0),(1,0,0))$. The height graph of this path is the second
one on Figure~\ref{fig:gr1}. We consider the edge expansion
$P(m,D,Q,s)$, with the data: $m=2$, $D=(0,-,1,\dots,1)$, $Q=(0,0)$, $s=1$.
\end{example}

\begin{figure}[hbt]
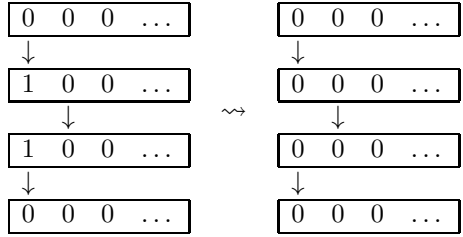

\begin{minipage}{3cm}
\[
\begin{array} {cccc}
\four{0}{0}{0}{\dots}
 \dara & \\
\four{1}{0}{0}{\dots}
 \darb & \\
\four{1}{0}{0}{\dots}
\dara& \\
\four{0}{0}{0}{\dots}
\end{array}
\]
\end{minipage}
$\rightsquigarrow$
\begin{minipage}{3cm}
\[
\begin{array} {cccc}
\four{0}{0}{0}{\dots}
 \dara & \\
\four{0}{0}{0}{\dots}
 \darb & \\
\four{0}{0}{0}{\dots}
 \dara & \\
\four{0}{0}{0}{\dots}
\end{array}
\]
\end{minipage}
\caption{An example of an exhaustive edge expansion.}
\label{table:example}
\end{figure}

The graphic presentation of this expansion is shown in
Figure~\ref{table:example}. The reader is invited to compare it
with~\cite[Figure 16(b)]{CR2}

We can see that this edge expansion is exhaustive. Indeed, 
we have 
\[A=\begin{array} {|c|c|c|c|c|}
\hline
1 & 0 & 0 & \dots & 0 \\ \hline
0 & 0 & 1 & \dots & 1 \\ \hline
\end{array},\]
and hence $|M(A)|=2$. Both of the $0$-monochromatic tuples $\alpha$
belong to the expansion, so the conditions of Definition~\ref{df:exh}
are satisfied. Also, we see that the obtained path $P(m,D,Q,s)$ is
admissible.

\section{The moves used by the algorithm}
\label{sect:4}

We now have all the technical tools needed to describe our algorithm, 
which in turn will be used to give a~new proof of 
Theorem~\ref{thm:cr2-paths}. Next, we proceed to define special cases
of the vertex and edge expansions.

\subsection{The summit move.} $\,$ 

\subsubsection{The definition.} $\,$

\vspace{5pt}

\noindent
Assume $P=(C,V)$ is an atomic path, and assume that we have
a~position~$i$, such that $2\leq i\leq k-1$, $h_i(P)>h_{i-1}(P)$, and
$h_i(P)>h_{i+1}(P)$. We call such an~index $i$ a~{\it summit} of~$P$. 
We cannot have $h_i(P)=1$, since that would imply $h_{i-1}(P)=h_{i+1}(P)=0$,
a~contradiction. Assume therefore that $h_i(P)\geq 2$. Since the path has 
even length, changing the orientation of the path will change the parity 
of $i$, so up to the above $\cs_2$-action we may assume that $i$ is odd.

Since we are allowed to apply $\cs_{[n]}$, we can furthermore assume
without loss of generality that $(C_{i-1},V_{i-1})=(0,1)$,
$(C_i,V_i)=(1,0)$, and $e_2^i=0$ (this is because $R^i(P)$ is not
$1$-monochromatic). In particular, we have 
$R^i(P)=(1,1,0,e_3^i,\dots,e_n^i)$.

With the data above, the {\it summit move on $P$ at position $i$} is
the vertex expansion $P(m,D,Q)$, where
\begin{itemize}
\item $m=i$;
\item $D=(0,0,0,\bar e_3^i,\bar e_4^i,\dots,\bar e_n^i)$;
\item $Q=(0,1,2,0,2,1)$.
\end{itemize}

\begin{figure}[hbt]
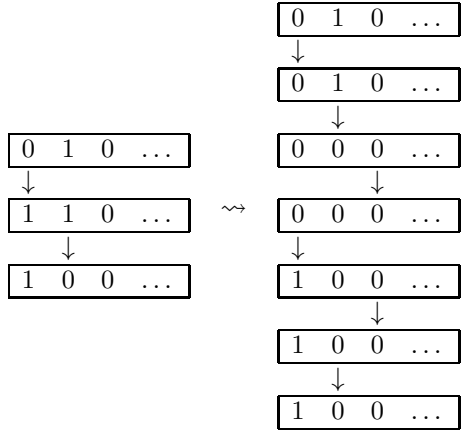

\begin{minipage}{3cm}
\[
\begin{array} {cccc}
\four{0}{1}{0}{\dots}
 \dara & \\
\four{1}{1}{0}{\dots}
 \darb & \\
\four{1}{0}{0}{\dots}
\end{array}
\]
\end{minipage}
$\rightsquigarrow$
\begin{minipage}{3cm}
\[
\begin{array} {cccc}
\four{0}{1}{0}{\dots}
 \dara & \\
\four{0}{1}{0}{\dots}
 \darb & \\
\four{0}{0}{0}{\dots}
 \darc & \\
\four{0}{0}{0}{\dots}
 \dara \\
\four{1}{0}{0}{\dots}
 \darc & \\
\four{1}{0}{0}{\dots}
 \darb & \\
\four{1}{0}{0}{\dots}
\end{array}
\]
\end{minipage}
\caption{The summit move.}
\label{table:summit}
\end{figure}

\begin{figure}[hbt]
\begin{center}

  \input{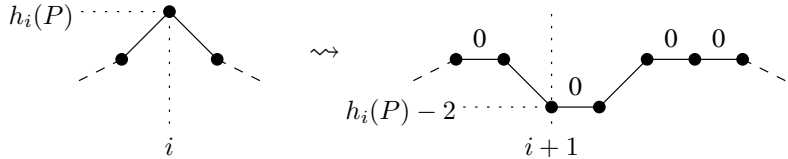}  

\end{center}
\caption{The height graph change in a summit move.}
\label{fig:summith}
\end{figure}

The summit move is illustrated by Figure~\ref{table:summit}, and 
the corresponding height graph change is shown in 
Figure~\ref{fig:summith}. We note that the vertex expansion discussed
in subsubsection~\ref{ssect:413} is an example of a~summit move.

\subsubsection{The summit move is exhaustive an yields an admissible 
path.}$\,$

\vskip5pt

\nin The following is the crucial propery of the summit move which
makes it useful for our algorithm.

\begin{prop} \label{prop:summ}
Assume that $P$ is an admissible path, and $P(m,D,Q)$ is the path
obtained from $P$ using a summit move. Then
\begin{enumerate}
\item[(1)] the vertex expansion $P(m,D,Q)$ is exhaustive;
\item[(2)] the path $P(m,D,Q)$ is admissible.
\end{enumerate}
\end{prop}
\pr We start by proving that $P(m,D,Q)$ is exhaustive. In the summit
move situation we have 
\[A=\begin{array} {|c|c|c|c|c|c|}
\hline
1 & 1 & 0 & e_3^i & \dots & e_n^i\\ \hline
0 & 0 & 0 & \bar e_3^i & \dots & \bar e_n^i\\ \hline
\end{array}\]
In particular, $M(A)$ consists of two $0$-monochromatic tuples, which
are neighbors in $\cuben$. If $h_i(P)=2$, then both of these
$0$-monochromatic tuples are contained in the path $P(m,D,Q)$ at
positions $i+1$ and $i+2$, see Figures~\ref{table:summit} 
and~\ref{fig:summith}. If $h_i(P)\geq 3$, then the path $P(m,D,Q)$
does not contain any of the two $0$-monochromatic tuples, but these
two tuples can be matched. In any case, the conditions of
Definition~\ref{df:exh} are satisfied, and $P(m,D,Q)$ is exhaustive.

Now we show that the path $P(m,D,Q)$ is admissible. Since $P$ is
atomic, we know that $k$ is even. By construction, the path $P(m,D,Q)$
has length $k+4$, so it is of even length as well. Furthermore, we
have $I(P(m,D,Q))=I(P)=0$, and $R^{k+4}(P(m,D,Q))=R^k(P)=0$. 

Clearly, since $P$ is atomic, the path $P(m,D,Q)$ does not contain any
$1$-monochromatic simplices. If $h_i(P)\geq 3$, then $P(m,D,Q)$ does
not contain any $0$-monochromatic simplices either, hence $P(m,D,Q)$
is atomic. Finally, assume $h_i(P)=2$. As mentioned above, in this
case the simplices $i+1$ and $i+2$ are $0$-monochromatic, so
$P(m,D,Q)$ is not atomic. However, since $i$ is odd, the subpaths
$P(m,D,Q)[1,i+1]$ and $P(m,D,Q)[i+2,k]$ are atomic, and the path
$P(m,D,Q)$ is a~concatenation of them. Thus, we conclude that in any
case the path $P(m,D,Q)$ is admissible.
\qed

\subsection{The plateau move.} 

\subsubsection{The definition.} $\,$

\vspace{5pt}

\noindent 
Assume again that $P=(C,V)$ is an~atomic path.  Assume we have
a~position~$i$, such that $2\leq i\leq k-2$,
$h_{i+1}(P)=h_i(P)>h_{i-1}(P)$. In such a~situation we say that we
have a~{\it plateau} at~$i$. If $h_i(P)=1$, then $h_{i-1}(P)=0$, 
contradicting the choice of~$i$. Thus we assume that $h_i(P)\geq 2$.

Since we are allowed to apply $\cs_{[n]}$, we can assume without loss
of generality that $(C_{i-1},V_{i-1})=(1,1)$, $(C_i,V_i)=(0,b)$, and
$(C_{i+1},V_{i+1})=(p,y)$, where $p\in[n]\setminus\{0\}$. As a~matter
of fact, again due to the $\cs_{[n]}$-action, we can assume that
either $p=1$ or $p=2$. These two cases lead to two slightly different
versions of the plateau move, which we consider separately.

\subsubsection{The generic plateau move.} $\,$

\vspace{5pt}

\noindent{\it The case:} $p=2$.  With the data above, the generic
plateau move on $P$ is the edge expansion $P(m,D,Q,s)$, where
\begin{itemize}
\item $m=i$;
\item $D=(-,0,e_2^i,e_3^i,\dots,e_n^i)$;
\item $Q=(1,2,1,2)$;
\item $s=2$.
\end{itemize}

\begin{figure}[hbt]
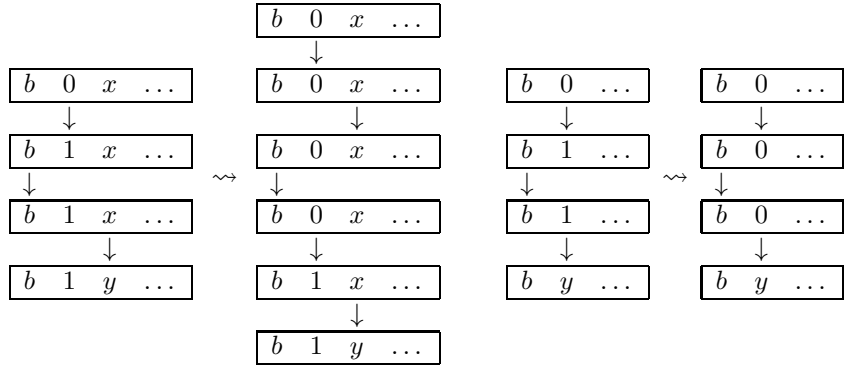

\begin{minipage}{2.7cm}
\[
\begin{array} {cccc}
\four{b}{0}{x}{\dots}
 \darb \\
\four{b}{1}{x}{\dots}
 \dara \\
\four{b}{1}{x}{\dots}
 \darc & \\
\four{b}{1}{y}{\dots}
\end{array}
\]
\end{minipage}
$\rightsquigarrow$
\begin{minipage}{2.7cm}
\[
\begin{array} {cccc}
\four{b}{0}{x}{\dots}
 \darb & \\
\four{b}{0}{x}{\dots}
 \darc & \\
\four{b}{0}{x}{\dots}
 \dara & \\
\four{b}{0}{x}{\dots}
 \darb \\
\four{b}{1}{x}{\dots}
 \darc & \\
\four{b}{1}{y}{\dots}
\end{array}
\]
\end{minipage}
\,\,\,\,\,\,\,\,
\begin{minipage}{2cm}
\[
\begin{array} {ccc}
\three{b}{0}{\dots}
 \darb \\
\three{b}{1}{\dots}
 \dara \\
\three{b}{1}{\dots}
 \darb \\
\three{b}{y}{\dots}
\end{array}
\]
\end{minipage}
$\rightsquigarrow$
\begin{minipage}{2cm}
\[
\begin{array} {ccc}
\three{b}{0}{\dots}
 \darb\\
\three{b}{0}{\dots}
 \dara \\
\three{b}{0}{\dots}
 \darb \\
\three{b}{y}{\dots}
\end{array}
\]
\end{minipage}
\vspace{0.4cm}
\caption{The generic and the special plateau moves.}
\label{table:plateau}
\end{figure}

\begin{figure}[hbt]
\begin{center}

  \input{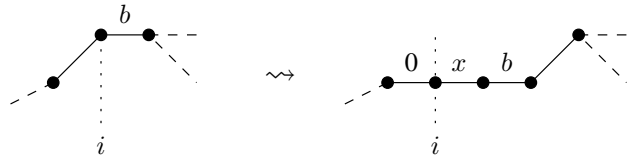}  

\end{center}
\caption{The height graph change in a generic plateau move.}
\label{fig:plath}
\end{figure}

The generic plateau move is shown on the left-hand side of
Figure~\ref{table:plateau}, and the corresponding height graph change
is shown in Figure~\ref{fig:plath}. Note, that the edge expansion
considered in the subsubsection~\ref{ssect:423} is an example of a~generic
plateau move.

\subsubsection{The special plateau move.} $\,$

\vspace{5pt}

\noindent{\it The case: $p=1$.}  
With the data above, the special plateau move on $P$ is the edge 
expansion $P(m,D,Q,s)$, where $m$ and $D$ are the same as in 
the generic case, and
\begin{itemize}
\item $Q=(1,1)$;
\item $s=1$.
\end{itemize}

The special plateau move is shown on the right hand side of
Table~\ref{table:plateau}, and the corresponding height graph change
is shown in Figure~\ref{fig:plath2}.

\subsubsection{Plateau moves are exhaustive and yield an admissible path.}$\,$

\vskip5pt

\nin Similarly to the case of the summit move, we have the following result.

\begin{figure}[hbt]
\begin{center}

  \input{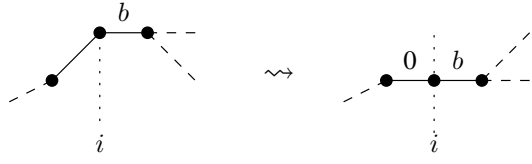}  

\end{center}
\caption{The height graph change in a special plateau move.}
\label{fig:plath2}
\end{figure}

\begin{prop} \label{prop:plm}
Assume that $P$ is an admissible path, and $P(m,D,Q,s)$ is the path
obtained from $P$ using a plateau move. Then
\begin{enumerate}
\item[(1)] the edge expansion $P(m,D,Q,s)$ is exhaustive;
\item[(2)] the path $P(m,D,Q,s)$ is admissible.
\end{enumerate}
\end{prop}
\pr We start by proving that $P(m,D,Q,s)$ is exhaustive. 
Both for generic and for special plateau moves we have 
\[A=\begin{array} {|c|c|c|c|c|c|}
\hline
b & 1 & e_2^i & e_3^i & \dots & e_n^i\\ \hline
b & 0 & e_2^i & e_3^i & \dots & e_n^i\\ \hline
\end{array}\]
In particular, $M(A)$ is empty, and hence $P(m,D,Q,s)$ is exhaustive,
unless $b=e_2^i=\dots=e_n^i$. So assume that $b=e_2^i=\dots=e_n^i$.
If $b=1$, then $R^i(P)$ is $1$-monochromatic, contradicting the
assumption that $P$ is atomic. If $b=0$, then $h_i(P)=1$,
contradicting the choice of $i$. We see that in any case, the edge
expansion $P(m,D,Q,s)$ is exhaustive.

Now we show that the path $P(m,D,Q,s)$ is admissible. Since $P$ is
atomic, we know that $k$ is even. By construction, the path
$P(m,D,Q,s)$ has length $k+2$ or $k$, so it is of even length as
well. Furthermore, we have $I(P(m,D,Q,s))=I(P)=0$, and
$R^{l}(P(m,D,Q,s))=R^k(P)=0$, where $l$ is the length of $P(m,D,Q,s)$.
Finally, it is clearly seen from Figures~\ref{fig:plath} and
\ref{fig:plath2} that the new path $P(m,D,Q,s)$ does not contain any
monochromatic simplices other than its end simplices, so it is
admissible.  \qed

\section{The main theorem}

\subsection{Reducibility of low admissible paths.}

\begin{df} 
A~path $P$ is called {\bf reducible} if there exists a sequence of
exhaustive vertex and edge expansions transforming it into a~$0$-path.
\end{df}


\begin{lm}\label{lm:h1}
Every admissible path $P$ such that $\max_i h_i(P)=1$ is reducible.
\end{lm}

\pr Clearly, it is enough to consider the case when $P=(C,V)$ is an
atomic path, such that $\max_i h_i(P)=1$. A~number of facts follows at
once from the assumption that $\max_i h_i(P)=1$. Specifically:
\begin{itemize}
\item we have $V_1=1$, $V_2=0$, $V_{k-2}=0$, $V_{k-1}=0$;
\item if $V_i=1$, for some $3\leq i\leq k-3$, then $V_{i-1}=V_{i+1}=0$;
\item after using $\cs_{[n]}$-symmetry we can assume that
  $R^2(P)=\dots=R^{k-1}(P)=(1,0,\dots,0)$.
\end{itemize}

To describe the reduction, we now describe a number of special vertex
and edge expansions. 

For the first two moves we assume that $3\leq i\leq k-3$ is chosen so
that $V_i=1$; which of course implies that $V_{i-1}=V_{i+1}=0$. Due to
$\cs_{[n]}$-action we may assume that $C_i=0$, $C_{i-1}=1$, and
$C_{i+1}\in\{1,2\}$.

\vspace{5pt}

\begin{figure}[hbt]
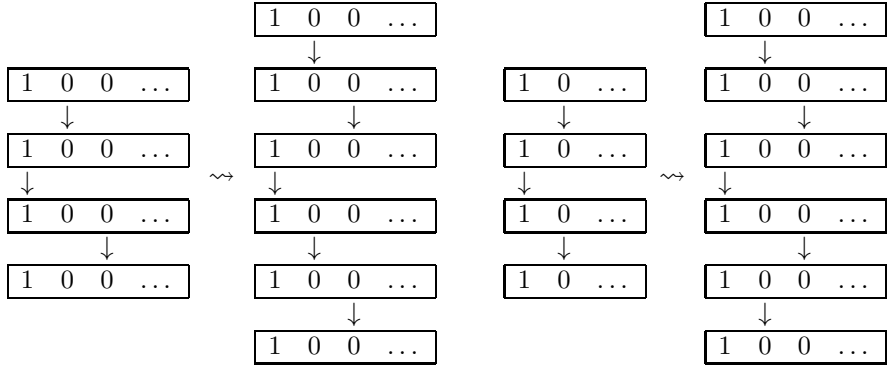

\begin{minipage}{2.7cm}
\[
\begin{array} {cccc}
\four{1}{0}{0}{\dots}
 \darb & \\
\four{1}{0}{0}{\dots}
 \dara & \\
\four{1}{0}{0}{\dots}
 \darc & \\
\four{1}{0}{0}{\dots}
\end{array}
\]
\end{minipage}
$\rightsquigarrow$
\begin{minipage}{2.7cm}
\[
\begin{array} {cccc}
\four{1}{0}{0}{\dots}
 \darb & \\
\four{1}{0}{0}{\dots}
 \darc & \\
\four{1}{0}{0}{\dots}
 \dara & \\
\four{1}{0}{0}{\dots}
 \darb \\
\four{1}{0}{0}{\dots}
 \darc & \\
\four{1}{0}{0}{\dots}
\end{array}
\]
\end{minipage}
\,\,\,\,\,\,\,\,
\begin{minipage}{2cm}
\[
\begin{array} {ccc}
\three{1}{0}{\dots}
 \darb  \\
\three{1}{0}{\dots}
 \dara  \\
\three{1}{0}{\dots}
 \darb \\
\three{1}{0}{\dots}
\end{array}
\]
\end{minipage}
$\rightsquigarrow$
\begin{minipage}{2.7cm}
\[
\begin{array} {cccc}
\four{1}{0}{0}{\dots}
 \darb & \\
\four{1}{0}{0}{\dots}
 \darc & \\
\four{1}{0}{0}{\dots}
 \dara & \\
\four{1}{0}{0}{\dots}
 \darc \\
\four{1}{0}{0}{\dots}
 \darb & \\
\four{1}{0}{0}{\dots}
\end{array}
\]
\end{minipage}
\caption{The moves which fatten a~unit: the left-hand side shows the
  case $C_{i-1}=2$, while the right-hand side shows the case
  $C_{i-1}=1$.}
\label{table:fatten}
\end{figure}

\begin{figure}[hbt]
\begin{center}

  \input{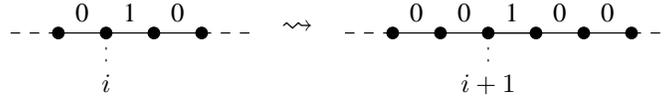}  

\end{center}
\caption{The height graph change when fattening units.}
\label{fig:fatten}
\end{figure}

\noindent 
{\bf Move 1: Fatten a unit.}
These are edge expansions of $P$ with the data
\begin{itemize}
\item $m=i$;
\item $D=(-,0,\dots,0)$;
\item $Q=(1,2,1,2)$ if $C_{i-1}=2$, $Q=(1,2,2,1)$ if $C_{i-1}=1$;
\item $s=2$.
\end{itemize}
This move is illustrated by Figure~\ref{table:fatten}, and the
corresponding height graph change is shown in Figure~\ref{fig:fatten}.

The resulting path $P(m,D,Q,s)$ has length $k+2$, which is even, and
Figure~\ref{fig:fatten} shows that this path is atomic. In this edge
expansion we have 
\[A=\begin{array} {|c|c|c|c|}
\hline
1 & 0 & \dots & 0 \\ \hline
1 & 0 & \dots & 0 \\ \hline
\end{array},\]
hence $M(A)$ is empty and the expansion is exhaustive.

\vspace{5pt}

\begin{figure}[hbt]
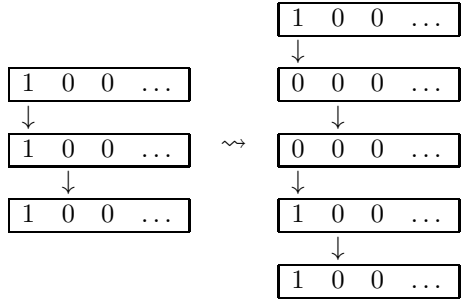

\begin{minipage}{3cm}
\[
\begin{array} {cccc}
\four{1}{0}{0}{\dots}
 \dara & \\
\four{1}{0}{0}{\dots}
 \darb & \\
\four{1}{0}{0}{\dots}
\end{array}
\]
\end{minipage}
$\rightsquigarrow$
\begin{minipage}{3cm}
\[
\begin{array} {cccc}
\four{1}{0}{0}{\dots}
 \dara & \\
\four{0}{0}{0}{\dots}
 \darb & \\
\four{0}{0}{0}{\dots}
 \dara & \\
\four{1}{0}{0}{\dots}
 \darb \\
\four{1}{0}{0}{\dots}
\end{array}
\]
\end{minipage}
\caption{The move eliminating a~unit.}
\label{table:elim}
\end{figure}

\begin{figure}[hbt]
\begin{center}

  \input{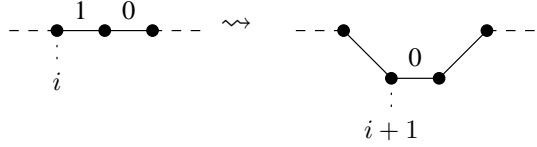}  

\end{center}
\caption{The height graph change when eliminating units.}
\label{fig:elim}
\end{figure}

\noindent
{\bf Move 2: Eliminate a unit.}  Assume that $i$ is odd. Eliminating
a~unit is a~vertex expansion of $P$ with the data
\begin{itemize}
\item $m=i$;
\item $D=(0,0,1,1,\dots,1)$;
\item $Q=(0,1,0,1)$.
\end{itemize}
This move is illustrated by Figure~\ref{table:elim}, and the
corresponding height graph change is shown in Figure~\ref{fig:elim}.

The resulting path $P(m,D,Q)$ has length $k+2$, which is even.
Figure~\ref{fig:elim} shows that this path is admissible, since $i$ is
odd: it is a~concatenation of atomic paths $P(m,D,Q)[1,i+1]$ and
$P(m,D,Q)[i+2,k]$. For this vertex expansion we have
\[A=\begin{array} {|c|c|c|c|c|}
\hline
1 & 0 & 0 & \dots & 0 \\ \hline
0 & 0 & 1 & \dots & 1 \\ \hline
\end{array},\]
hence $|M(A)|=2$. Both $0$-monochromatic tuples belong to the
expansion, so it is exhaustive.

\vspace{5pt}

\begin{figure}[hbt]
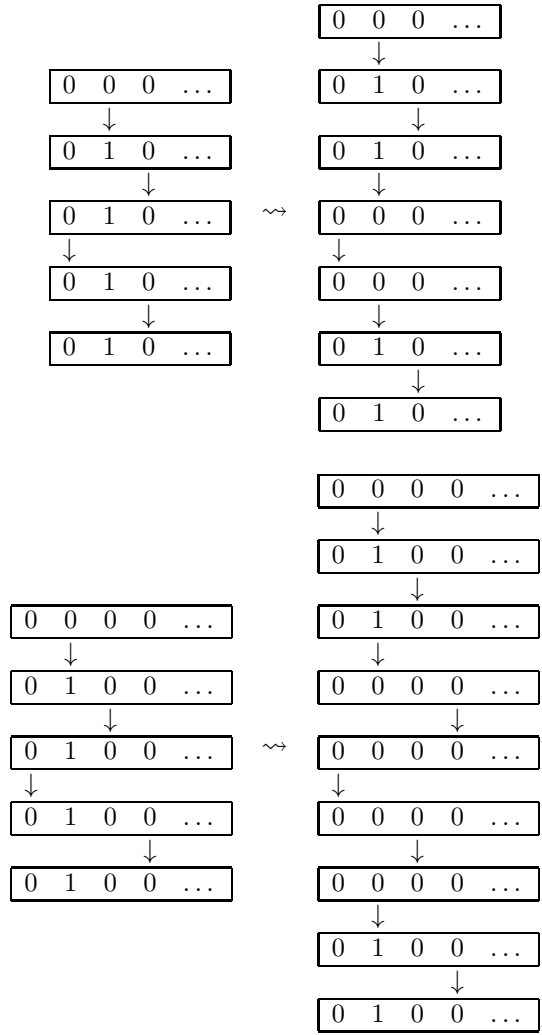

\begin{minipage}{3cm}
\[
\begin{array} {cccc}
\four{0}{0}{0}{\dots}
 \darb & \\
\four{0}{1}{0}{\dots}
 \darc & \\
\four{0}{1}{0}{\dots}
 \dara & \\
\four{0}{1}{0}{\dots}
 \darc & \\
\four{0}{1}{0}{\dots}
\end{array}
\]
\end{minipage}
$\rightsquigarrow$
\begin{minipage}{3cm}
\[
\begin{array} {cccc}
\four{0}{0}{0}{\dots}
 \darb & \\
\four{0}{1}{0}{\dots}
 \darc & \\
\four{0}{1}{0}{\dots}
 \darb & \\
\four{0}{0}{0}{\dots}
 \dara \\
\four{0}{0}{0}{\dots}
 \darb & \\
\four{0}{1}{0}{\dots}
 \darc & \\
\four{0}{1}{0}{\dots}
\end{array}
\]
\end{minipage}
\vskip10pt
\begin{minipage}{3.5cm}
\[
\begin{array} {ccccc}
\five{0}{0}{0}{0}{\dots}
 \darb && \\
\five{0}{1}{0}{0}{\dots}
 \darc && \\
\five{0}{1}{0}{0}{\dots}
 \dara && \\
\five{0}{1}{0}{0}{\dots}
 &&&\dar & \\
\five{0}{1}{0}{0}{\dots}
\end{array}
\]
\end{minipage}
$\rightsquigarrow$
\begin{minipage}{3.5cm}
\[
\begin{array} {ccccc}
\five{0}{0}{0}{0}{\dots}
 \darb & \\
\five{0}{1}{0}{0}{\dots}
 \darc & \\
\five{0}{1}{0}{0}{\dots}
 \darb & \\
\five{0}{0}{0}{0}{\dots}
&&& \dar \\
\five{0}{0}{0}{0}{\dots}
 \dara & \\
\five{0}{0}{0}{0}{\dots}
 \darc & \\
\five{0}{0}{0}{0}{\dots}
 \darb & \\
\five{0}{1}{0}{0}{\dots}
 &&& \dar & \\
\five{0}{1}{0}{0}{\dots}
\end{array}
\]
\end{minipage}
\caption{The moves shortening zeroes.}
\label{table:short}
\end{figure}

\begin{figure}[hbt]
\begin{center}

  \input{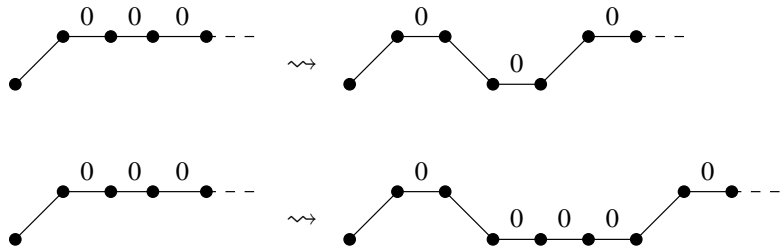}  

\end{center}
\caption{The height graph changes when shortening zeroes.}
\label{fig:short}
\end{figure}

\noindent 
{\bf Move 3: Shorten zeroes.}  For this move we assume that
$V_2=V_3=V_4=0$. Furthermore, due to $\cs_{[n]}$-action, we may assume
that $C_1=1$, $C_2=2$, $C_3=0$, and $C_4\in\{2,3\}$. The move is an
edge expansion of $P$ with the data
\begin{itemize}
\item $m=3$;
\item $D=(-,0,0,1,1,\dots,1)$ if $C_4=2$; $D=(-,0,0,0,1,1,\dots,1)$ if
  $C_4=3$;
\item $Q=(2,1,1,2)$ if $C_4=2$; $Q=(2,1,3,2,1,3)$ if $C_4=3$;
\item $s=2$ if $C_4=2$; $s=3$ if $C_4=3$.
\end{itemize}
This move is illustrated by Figure~\ref{table:short}, and the 
corresponding height graph changes are shown in Figure~\ref{fig:short}.

The resulting path $P(m,D,Q,s)$ has length $k+2$ or $k+4$, which is
even.  Figure~\ref{fig:short} shows that this path is admissible: it
is a~concatenation of either $2$ or $3$ atomic paths. Furthermore, if
$C_4=2$, we have
\[A=\begin{array} {|c|c|c|c|c|c|}
\hline
0 & 1 & 0 & 0 & \dots & 0 \\ \hline
0 & 0 & 0 & 1 & \dots & 1 \\ \hline
\end{array}.\]
We have $|M(A)|=4$. Two of the $0$-monochromatic tuples belong to the
expansion, and the other two can be matched to each other, so the
expansion is exhaustive.  If $C_4=3$, we have
\[ 
A=\begin{array} {|c|c|c|c|c|c|c|}
\hline
0 & 1 & 0 & 0 & 0 & \dots & 0 \\ \hline
0 & 0 & 0 & 0 & 1 & \dots & 1 \\ \hline
\end{array},\]
hence $|M(A)|=8$. One can see that $4$ of the $0$-monochromatic tuples
belong to the expansion, and there is a~complete matching on the
remaining $4$. Hence again the expansion is exhaustive.  

\vspace{5pt}

We can now use these $3$ types of moves to reduce an arbitrary atomic
path $P$, such that $\max_i h_i(P)=1$, to a~$0$-path. To start with, 
we can apply Move $1$ to all even $i$, such that $V_i=1$. When
applying such a~move we change the parity of the $1$-edge at position
$i$, see Figure~\ref{fig:fatten}, while all other $1$-edges keep the
same parity: those before $i$ have the same index, and those after $i$
will be shifted by~$2$. When we are done, all $1$-edges start at odd
positions. At this point, we can eliminate them one-by-one using
Move~$2$. We will be left with having to consider the case of the
atomic path $P$ such that $V_2=V_3=\dots=V_k=0$. Now using Move~$3$,
we can reduce the length of $P$, until $k=4$, and this case was dealt
with in the Example~\ref{ex:p4}.
\qed


\subsection{Reducibility of all admissible paths.} $\,$

\vskip5pt

\noindent
We are now ready to formulate our main theorem.

\begin{thm}\label{thm:main}
Every admissible path is reducible.
\end{thm}



\pr Clearly, it is enough to consider the case when $P$ is
atomic. Assume that the statement of the theorem is false, and pick an
atomic, but not reducible path~$P$ of length~$k$, such that $H(P)$ is
minimal with respect to the reverse lexicographic order. By
Lemma~\ref{lm:h1} we can assume that $\max_i h_i(P)\geq 2$.

Assume first, that the path $P$ has a~summit $i$, $2\leq i\leq
k-1$. By reversing the direction of the path, if needed, we may assume
that $i$ is odd. Let $\wti P$ be the path resulting from $P$ by
applying the summit move at position~$i$. By
Proposition~\ref{prop:plm}, that vertex expansion is exhaustive, and
the obtained path is admissible.

If $h_i(P)\geq 3$, the path $\wti P$ is in fact atomic. On the other
hand, as is visible from Figure~\ref{fig:summith}, we have $H(P)\succ
H(\wti P)$, so by the choice of $P$, the path $\wti P$ must be
reducible. This implies that $P$ is reducible as well.

If $h_i(P)=2$, the obtained path $\wti P$ is not atomic, since
$R^{i+1}(\wti P)=R^{i+2}(\wti P)$ are $0$-mono\-chromatic. However,
both subpaths $\wti P_1=\wti P[1,i+1]$ and $\wti P_2=\wti P[i+2,k]$
are atomic. Since $H(P)\succ H(\wti P_1)$ and $H(P)\succ H(\wti P_2)$,
both paths $\wti P_1$ and $\wti P_2$ are reducible, which implies that
$P$ is reducible as well.

We can thus assume that $P$ has no summits. In this case, we pick $i$,
$2\leq i\leq k-1$, to be the minimal index such that $h_i(P)=\max_j
h_j(P)\geq 2$. In particular, we have $h_i(P)>h_{i-1}(P)$, which,
since $P$ has no summits, implies $h_i(P)=h_{i+1}(P)$. Thus, the path
$P$ has a~plateau at~$i$, and we let $\wti P$ denote the path obtained
from $P$ by applying the corresponding plateau move. Here, we use the
kind of plateau move which is appropriate to the situation, i.e., we
use the specific plateau move if $C_{i-1}=C_{i+1}$, and we use the
generic plateau move otherwise. In any case, by
Proposition~\ref{prop:summ}, this edge expansion is exhaustive, and
the resulting path is atomic. On the other hand, we have $H(P)\succ
H(\wti P)$, implying by the choice of $P$, that the path $\wti P$ is
reducible. This implies that the path $P$ is reducible as well,
leading to the contradiction to our original assumption.  \qed


\subsection{Subdividing geometric simplex paths} $\,$

\noindent
Having gained the understanding of how the vertex and edge expansions
can be interpreted geometrically, we are now ready to apply our
combinatorial theory to the geometric simplicial context. In fact, we
derive Theorem~\ref{thm:cr2-paths} as a~straighforward corollary of
Theorem~\ref{thm:main}.

\vskip5pt

\noindent
{\bf Proof of Theorem~\ref{thm:cr2-paths}.}  
Assume we are given an~$n$-dimensional geometric simplex path $\Sigma$ 
in the standard form, and let $P$ be the associated abstract simplex 
path. Since $\Sigma$ is in standard form, the path $P$ is atomic. 
Theorem~\ref{thm:main} states that there is a~sequence of exhaustive 
vertex and edge expansions transforming the abstract simplex path~$P$ 
into an~abstract simplex path~$\widetilde P$, which is a~$0$-path of even
length. It follows from our geometric interpretation of these expansions, 
that we obtain a~subdivision $\widetilde S$ of $\Div(\da^n)$, and 
an~extension of the binary labeling $b$ to $\widetilde S$, such that
\begin{enumerate}
\item[(1)] only the interior of $\Sigma$ is subdivided;
\item[(2)] all monochromatic simplices in $\widetilde S$ are
  $0$-monochromatic, some of them are matched in pairs, each pair
  having a~common $(n-1)$-simplex, and the rest forms a~geometric
  simplex path $\widetilde\Sigma$, which corresponds to the abstract
  simplex path~$\widetilde P$, which is a~$0$-path of even length.
\end{enumerate}
Since $\widetilde P$ has even length, also the $0$-monochromatic 
simplices forming $\widetilde\Sigma$ can be broken in matched pairs, 
with $n$-simplices in each pair sharing an~$(n-1)$-simplex.

It is now a~simple fact, that two $0$-monochromatic $n$-simplices
$\sigma_1$ and $\sigma_2$, which are sharing an~$(n-1)$-simplex
$\sigma_1\cap\sigma_2$, can be further subdivided to eliminate all
monochromatic simplices, such that only the interior of $\sigma_1$,
$\sigma_2$, and $\sigma_1\cap\sigma_2$ are subdivided. To do that,
simply subdivide $\sigma_1\cap\sigma_2$ as a~Schlegel diagram, cone
over it with the apexes at $\sigma_1\setminus(\sigma_1\cap\sigma_2)$
and $\sigma_2\setminus(\sigma_1\cap\sigma_2)$, and take all the new
labels to be~$1$. One may think of it as an edge expansion with
$D=(1,\dots,1)$ without specifying $m$ and~$Q$. This is the same
subdivision as in~\cite[Figure~15]{CR2}.

This further subdivision of $\widetilde S$ leads to the subdivision
$S(\Div(\da^n))$, which has the desired properties. \qed

\section{Appendix}

\subsection{Simplified summit move}$\,$

\vskip5pt

\nin
One of the purposes of our algorithm in Section~\ref{sect:4} was to
demonstrate, that one can expand any atomic path to a~path $\widetilde
P$, which has a~simple form, namely $\max_i h_i(\widetilde P)\leq 1$,
using few types of moves: the summit move and the plateau moves. If
instead the focus is on having as simple moves as possible, instead of
as few as possible, it might be beneficial to consider the following
additional move, which we call \emph{simplified summit move}.

The simplified summit move on a~path $P$ at position $i$ is
the vertex expansion $P(m,D,Q)$, where
\begin{itemize}
\item $m=i$;
\item $D=(0,0,e_2^i,e_3^i,\dots,e_n^i)$;
\item $Q=(0,1,0,1)$.
\end{itemize}

\begin{figure}[hbt]
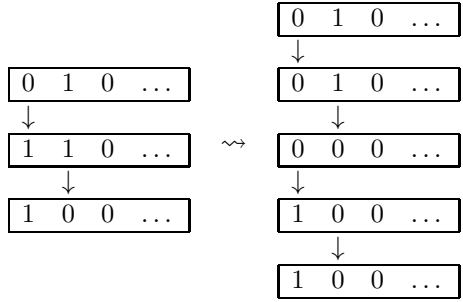

\begin{minipage}{3cm}
\[
\begin{array} {cccc}
\four{0}{1}{0}{\dots}
 \dara & \\
\four{1}{1}{0}{\dots}
 \darb & \\
\four{1}{0}{0}{\dots}
\end{array}
\]
\end{minipage}
$\rightsquigarrow$
\begin{minipage}{3cm}
\[
\begin{array} {cccc}
\four{0}{1}{0}{\dots}
 \dara & \\
\four{0}{1}{0}{\dots}
 \darb & \\
\four{0}{0}{0}{\dots}
 \dara \\
\four{1}{0}{0}{\dots}
 \darb & \\
\four{1}{0}{0}{\dots}
\end{array}
\]
\end{minipage}
\caption{The simplified summit move.}
\label{table:summit-s}
\end{figure}

\begin{figure}[hbt]
\begin{center}

  \input{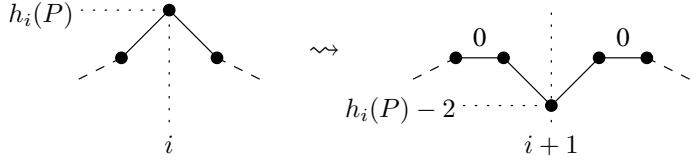}  

\end{center}
\caption{The height graph change in a~simplified summit move.}
\label{fig:summith-s}
\end{figure}

The simplified summit move is shown on Figure~\ref{table:summit-s}, and
the corresponding height graph change is shown in
Figure~\ref{fig:summith-s}.

\begin{prop} \label{prop:summ-s}
Assume that $P$ is an admissible path, with summit at $i$ such that
$h_i(P)\geq 3$, and $P(m,D,Q)$ is the path obtained from
$P$ using a simplified summit move at position~$i$. Then
\begin{enumerate}
\item[(1)] the vertex expansion $P(m,D,Q)$ is exhaustive;
\item[(2)] the path $P(m,D,Q)$ is admissible.
\end{enumerate}
\end{prop}
\pr The proof is basically the same as that of
Proposition~\ref{prop:summ}.  The only difference is that here we have
\[A=\begin{array} {|c|c|c|c|c|c|}
\hline
1 & 1 & e_2^i & e_3^i & \dots & e_n^i\\ \hline
0 & 0 & e_2^i & e_3^i & \dots & e_n^i\\ \hline
\end{array}.\]
In the simplified summit move situation, the set $M(A)$ is empty,
there are no monochromatic simplices: at least one of the labels
$e_2^i,\dots,e_n^i$ is equal to $0$, since otherwise $P$ would have
a~$1$-monochromatic simplex, and at least one of them is equal to~$1$,
since $h_i(P)\geq 3$. We conclude that the vertex expansion $P(m,D,Q)$
is exhaustive. It is admissible again due to the assumption
$h_i(P)\geq 3$.  \qed

\vspace{5pt}

\noindent
The simplified summit move can be used in place of the regular summit
move at position~$i$, whenever $h_i(P)\geq 3$. It is only in the case
$h_i(P)=2$ that we need to resort to the full summit move, as otherwise
the obtained path would not be admissible.

\subsection{Examples}$\,$

\vskip5pt

\nin
To have a collection of examples, let us analyze all atomic paths of
length $k=2,4,6$. We see, that up to the action of the symmetry group
$\cs_{[n]}$, there is only one atomic path of length $2$, namely
$P=((0),(0))$, and there is only one atomic path of length $4$, namely
$P=((0,1,0),(1,0,0))$.  The numbers of atomic paths of length $6$ are
shown in Figure~\ref{table:p6}.

\begin{figure}[hbt]
\[\begin{array}{|c|c|c|}
\hline
\textrm{value of }n & \sharp\textrm{ atomic paths } &
\sharp\textrm{ atomic paths mod reflection }\\ \hline
n=1 & 2 & 2 \\ \hline
n=2 & 8 & 7\\ \hline
n\geq 3 & 9 & 8 \\
\hline
\end{array}\]
\vspace{0.1cm}
\caption{Number of atomic paths of length~$6$.}
\label{table:p6}
\end{figure}
\begin{figure}[hbt]
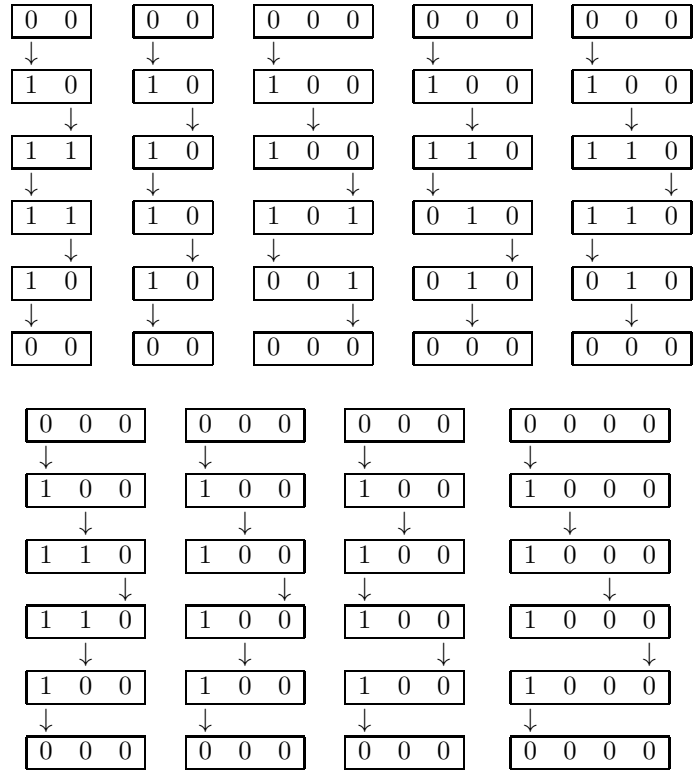

\begin{minipage}{1.5cm}
\[
\begin{array} {cc}
\two{0}{0} \dar& \\ \two{1}{0} &\dar \\ \two{1}{1} \dar& \\ \two{1}{1}
&\dar \\ \two{1}{0} \dar& \\ \two{0}{0}
\end{array}
\]
\end{minipage}
\begin{minipage}{1.5cm}
\[
\begin{array} {cc}
\two{0}{0}
 \dar&  \\
\two{1}{0}
 &\dar \\
\two{1}{0}
 \dar& \\
\two{1}{0}
 &\dar \\
\two{1}{0}
 \dar& \\
\two{0}{0}
\end{array}
\]
\end{minipage}
\begin{minipage}{2cm}
\[
\begin{array} {ccc}
\three{0}{0}{0}
 \dara  \\
\three{1}{0}{0} 
 \darb \\
\three{1}{0}{0}
 \darc \\
\three{1}{0}{1}
\dara \\
\three{0}{0}{1}
 \darc \\
\three{0}{0}{0}
\end{array}
\]
\end{minipage}
\begin{minipage}{2cm}
\[
\begin{array} {ccc}
\three{0}{0}{0}
 \dara  \\
\three{1}{0}{0}
 \darb \\
\three{1}{1}{0}
 \dara \\
\three{0}{1}{0}
 \darc \\
\three{0}{1}{0}
 \darb \\
\three{0}{0}{0}
\end{array}
\]
\end{minipage}
\begin{minipage}{2cm}
\[
\begin{array} {ccc}
\three{0}{0}{0}
 \dara  \\
\three{1}{0}{0}
 \darb \\
\three{1}{1}{0}
 \darc \\
\three{1}{1}{0}
 \dara \\
\three{0}{1}{0}
 \darb \\
\three{0}{0}{0}
\end{array}
\]
\end{minipage}
\vskip10pt
\begin{minipage}{2cm}
\[
\begin{array} {ccc}
\three{0}{0}{0}
 \dara  \\
\three{1}{0}{0}
 \darb \\
\three{1}{1}{0}
 \darc \\
\three{1}{1}{0}
 \darb \\
\three{1}{0}{0}
 \dara \\
\three{0}{0}{0}
\end{array}
\]
\end{minipage}
\begin{minipage}{2cm}
\[
\begin{array} {ccc}
\three{0}{0}{0}
 \dara  \\
\three{1}{0}{0}
 \darb \\
\three{1}{0}{0}
 \darc \\
\three{1}{0}{0}
 \darb \\
\three{1}{0}{0}
 \dara \\
\three{0}{0}{0}
\end{array}
\]
\end{minipage}
\begin{minipage}{2cm}
\[
\begin{array} {ccc}
\three{0}{0}{0}
 \dara  \\
\three{1}{0}{0}
 \darb \\
\three{1}{0}{0}
 \dara \\
\three{1}{0}{0}
 \darc \\
\three{1}{0}{0}
 \dara \\
\three{0}{0}{0}
\end{array}
\]
\end{minipage}
\begin{minipage}{2.7cm}
\[
\begin{array} {cccc}
\four{0}{0}{0}{0}
 \dara & \\
\four{1}{0}{0}{0}
 \darb & \\
\four{1}{0}{0}{0}
 \darc & \\
\four{1}{0}{0}{0}
 \dard \\
\four{1}{0}{0}{0}
 \dara & \\
\four{0}{0}{0}{0}
\end{array}
\]
\end{minipage}
\caption{Atomic paths of length $6$.}
\label{table:p6d}
\end{figure}

The representatives of all equivalence classes of atomic paths of
length~$6$ under the $S_{[n]}$-action are listed in the
Figure~\ref{table:p6d}.  If, in addition, the reflection action is
taken into account, the paths $3$ and $4$ in that table would be
identified as well.

\end{document}